\documentclass{aa}  
\usepackage{graphicx}
\usepackage{txfonts}

\usepackage{epstopdf}

\usepackage{amsmath}

\usepackage{enumitem}
\usepackage{natbib}
\bibpunct{(}{)}{;}{a}{}{,}

\usepackage{multirow}

\begin{document}

   \title{OMC-2 FIR\,4 under the microscope: Shocks, filaments, and a highly collimated jet at 100 au scales}

   \author{L. Chahine
         \inst{1,2}  
          \and
    A. L\'{o}pez-Sepulcre\inst{1,3} \and
    L. Podio\inst{4} \and
   C. Codella\inst{4,3} \and
         R. Neri\inst{1} \and
     S. Mercimek \inst{4,5} \and
     M. De Simone\inst{3,4} \and
     P. Caselli\inst{6} \and
         C. Ceccarelli\inst{3} \and
        M. Bouvier\inst{3} \and
        N. Sakai\inst{7} \and
        F. Fontani\inst{3} \and
          S. Yamamoto\inst{8} \and
        F. O. Alves \inst{6} \and
        V. Lattanzi\inst{6} \and
          L. Evans\inst{9,4}  \and
           C. Favre \inst{3} }

   \institute{Institut de Radioastronomie Millim\'etrique (IRAM), 300 rue de la Piscine, 38406 Saint-Martin-d'H\`eres, France \\ e-mail: chahine@iram.fr\
         \and
              Universit\'e Grenoble Alpes, 38000 Grenoble, France \ 
             \and Universit\'e Grenoble Alpes, CNRS, IPAG, 38000 Grenoble, France \
            \and INAF, Osservatorio Astrofisico di Arcetri, Largo E. Fermi 5, I-50125, Firenze, Italy \
            \and Universit\`{a} degli Studi di Firenze, Dipartimento di Fisica e Astronomia, Via G. Sansone 1, 50019 Sesto Fiorentino, Italy \
            \and Center for Astrochemical Studies, Max Planck Institute for Extraterrestrial Physics, Garching, 85748, Germany \
            \and
           The Institute of Physical and Chemical Research (RIKEN), Saitama 351-0198, Japan \
            \and
            Department of Physics, The University of Tokyo, Bunkyo-ku, Tokyo 113-0033, Japan \
            \and  IRAP, Université de Toulouse, 9 avenue du colonel Roche, 31028 Toulouse Cedex 4, France}

   \date{Received day month year; Accepted ...}

 
  \abstract 
   {Star-forming molecular clouds are characterised by the ubiquity of intertwined filaments. The filaments have been observed in both high- and low-mass star-forming regions, and they are thought to split into collections of sonic fibres.  
   The locations where filaments converge are termed hubs, and these are associated with the young stellar clusters.
   However, the observations of filamentary structures within hubs at distances of 75-300 pc require a high angular resolution $<$2$\arcsec$ ($\sim$ 150-600 au) that limits the number of such studies conducted so far.}
   {The integral shaped filament (ISF) of the Orion A molecular cloud is noted for harbouring several hubs within which no filamentary structures have been observed so far. The goal of our study is to investigate the nature of the filamentary structures within one of these hubs, which is the chemically rich hub OMC-2\,FIR\,4, and to analyse their emission with high density and shock tracers.}
   {We observed the OMC-2\,FIR\,4 proto-cluster using Band 6 of the Atacama Large (sub-)Millimetre Array (ALMA) in Cycle 4 with an angular resolution of $\sim$0.26$\arcsec$ (100 au). We analysed the spatial distribution of dust, the shock tracer SiO, and dense gas tracers (i.e., CH$_{3}$OH, CS, and H$^{13}$CN). We also studied the gas kinematics using SiO and CH$_{3}$OH maps.}
   {Our observations for the first time reveal interwoven filamentary structures within OMC-2\,FIR\,4 that are probed by several tracers. Each filamentary structure is characterised by a distinct velocity as seen from the emission peak of CH$_{3}$OH lines. They also show transonic and supersonic motions. SiO is associated with filaments and also with multiple bow-shock features. The bow-shock features have sizes between $\sim$500 and 2700 au and are likely produced by the outflow from HOPS-370. Their dynamical ages are $<$800 yr. In addition, for the first time, we reveal a highly collimated SiO jet ($\sim$1$^{\circ}$) with a projected length of $\sim$5200 au from the embedded protostar VLA15.} 
   {Our study unveiled the previously unresolved filamentary structures as well as the shocks within OMC-2\,FIR\,4. The kinematics of the filamentary structures might be altered by external and/or internal mechanisms such as the wind from H\,II regions, the precessing jet from the protostellar source HOPS-370, or the jet from VLA\,15. While the complexity of the region, coupled with the limited number of molecular lines in our dataset, makes any clear association with these mechanisms challenging, our study shows that multi-scale observations of these regions are crucial for understanding the accretion processes and flow of material that shape star formation.}


   \maketitle
%

\section{Introduction}

\label{Intro}

The interstellar medium (ISM) is highly filamentary on all scales. Networks of intricate filaments are seen in low- and high-mass star-forming regions, and even in clouds where no star formation is yet ongoing (e.g. \citealt{Goldsmith2008,Andre2014, Molinari2014, Orkisz2019, Chung2021}). In recent years, high angular resolution studies have shown that the filaments split into small-scale and velocity-coherent elongated structures, called fibres, which in turn fragment into the dense cores we observe today (e.g. \citealt{Hacar2013,Hacar2017-ngc,Tafalla2015, Henshaw2017, Sokolov2019}). Among star-forming regions, the Orion A cloud is noted for its prominent filamentary structures in both CO and dust maps (e.g. \citealt{Bally1987, Chini1997, Johnstone1999, Shimajiri2011, Polychroni2013, Andre2014, Suri2019}). Within this cloud lies the well-known integral shaped filament (ISF; \citealt{Bally1987}), the most massive filament in the solar neighbourhood with a total mass per unit length (M$_{lin} \sim$500 M$_{\odot}$\,pc$^{-1}$; \citealt{Bally1987}). The ISF is located at a distance of $\sim$393$\pm$25 pc \citep{Grossschedl2018}. Several studies have found that the ISF consists of substructures of small-scale ($<$1 pc) sub-filaments or fibres (e.g. \citealt{Martin-Pintado1990, Rodriguez-Franco1992, Shimajiri2011, Li2013, Hacar2017-ori, Hacar2018, Zhang2020}) and condensations or hub-filament structures (hubs) \citep{Mezger1990, Chini1997, Myers2009, Shimajiri2011, Hacar2018}. In turn, the latter are thought to contain complex nets of fibres \citep{Hacar2018} or of filamentary structures. However, the filamentary structures within the hubs along the ISF were not resolved at 4$\farcs$5 resolution ($\sim$ 2000 au; \citealt{Hacar2018}). To our knowledge, filamentary structures within a hub were only observed within IRAS4A at 300 pc. They have sizes $\leq$ 450 au \citep{DeSimone2022}.\ To observe similar structures within the hubs of the ISF, an angular resolution of about 2$\arcsec$ or smaller ($<$ 600 au) is therefore required. Because the filament and fibre nomenclature refers to structures at larger scales, we decided to adopt the name "filamentary structures" for those at $<$1000 au scales.
\newline \indent
In this work, we focus on the intermediate-mass hub OMC-2\,FIR\,4, an archetypical Sun-like star-forming region along the ISF. This region was identified for the first time as a bright continuum source at 1.3 mm by \cite{Mezger1990}. Later, it was observed, studied, and characterised at several wavelengths. Based on its envelope mass and luminosity ($\sim$30 M$_{\odot}$ and $\sim$400 L$_{\odot}$; \citealt{Mezger1990}), OMC-2\,FIR\,4 was considered as an intermediate-mass protostar \citep{Johnstone2003}. However, \cite{Shimajiri2008} discovered that OMC-2\,FIR\,4 is actually a protostellar cluster hosting several dusty cores and not a single protostar. Since then, OMC-2\,FIR\,4 was thoroughly studied with different telescopes and at different wavelengths (e.g. \citealt{Takahashi2008, Kama2013, Kama2015, Furlan2014, Lopez-Sepulcre2017, Evans2022}). Its envelope size and luminosity were estimated to be $\sim$10$^{4}$\,au and $\leq$1000 L$_{\odot}$ \citep{Crimier2009, Furlan2014}. Its systemic velocity is $\sim$11.4 km\,s$^{-1}$ \citep{Shimajiri2008,Favre2018}. The clustered nature of OMC-2\,FIR\,4 was further confirmed by several studies \citep{Lopez-Sepulcre-and-taquet2013, Kainulainen2017, Tobin2019, Chahine2022}. In addition, more of its physical properties have been revealed. Its outer shell is strongly irradiated by a far-ultraviolet (FUV) field ($\sim$1500 $G_{\rm 0}$,  where $G_{0}$ is the FUV radiation field in Habing units), arising from the high-mass stars located in the nearby Trapezium OB association \citep{Loepz-Sepulcre-and-kama2013, Favre2018}, and its interior envelope is subject to high irradiation and ionisation by local cosmic ray-like particles with an ionisation rate of $\zeta \sim$ 4 $\times$ 10$^{-14}$ s$^{-1}$ \citep{Ceccarelli2014, Fontani2017, Favre2018}. Due to these characteristics, OMC-2\,FIR\,4 is considered as the nearest analogue of our solar birth environment, which is thought to be a clustered dense environment \citep{Adams2010,Pfalzner2015} that was subject to internal irradiation from energetic particles (>10 MeV; \citealt{Gounelle2013}).
\newline
Despite all this progress, several questions still swirl around OMC-2\,FIR\,4. The observations at parsec and sub-parsec scales using quiescent gas tracers (e.g. N$_{2}$H$^{+}$; \citealt{Hacar2018,Fontani2020}, NH$_{3}$; \citealt{Wu2018}, HC$_{3}$N; \citealt{Tatematsu2008}) have shown that OMC-2\,FIR\,4 is a hub-filament structure. In this context, the open questions are whether filamentary structures exist within this hub, and if they might be observed at protocluster scale ($\leq$ 0.1 pc) using dynamic gas tracers (such as SiO and CH$_{3}$OH), as in the case of NGC 1333 IRAS\,4A \citep{DeSimone2022}. The differences that might arise from this type of observations compared to the studies performed at large scales are unclear. The same applies to the differences between the dynamic gas (traced by, e.g., SiO and CH$_{3}$OH) and the quiescent gas (traced by, e.g., N$_{2}$H$^{+}$ and NH$_{3}$).
Another important question we also address is the presence of jets and outflows. It is known that within the protocluster, the hot corino HOPS-108 drives a jet towards the south-east \citep{Lattanzi2022}. In addition, OMC-2\,FIR\,4 is located to the south of the intermediate-mass protostar HOPS-370 that drives an outflow whose southern lobe overlaps, at least in projection, with OMC-2\,FIR\,4 \citep{Shimajiri2008}. The outflow was detected in several tracers such as CO and [O\,I] (e.g. \citealt{Williams2003, Shimajiri2008, Takahashi2008, Gonzalez-Garcia2016, Tobin2019}), and several non-thermal jet knots were also discovered \citep{Osorio2017}. However, whether it has triggered the star formation in OMC-2\,FIR\,4 is still an open question.
\newline \indent
Two key molecules that can help to answer these questions and to study the diverse processes that occur in molecular clouds and stellar protoclusters are SiO and CH$_{3}$OH. First, SiO is a useful tracer of shocked gas and outflows. During shock processes, gas-grain collisions (sputtering) and grain-grain collisions (shattering) release Si from the grains into the gas phase \citep{Draine1983, Flower1994, Schilke1997, Caselli1997}. Si is oxidised rapidly in SiO, resulting in an increase of SiO abundance in the shocks by several orders of magnitude \citep{Gusdorf2008a, Gusdorf2008b, Guillet2011}. On the other hand, CH$_{3}$OH is present in different environments such as molecular clouds, photo-dissociation regions (PDR), and hot corinos (e.g. \citealt{Leurini2004, Leurini2010,Cuadrado2017,Maret2005, Bouvier2020}). It is prevalently formed on dust grains (e.g. \citealt{Tielens1982, Watanabe2002, Rimola2014}) and is released into the gas phase via thermal and/or non-thermal processes  (e.g. \citealt{Duley1993, Flower1994, Minissale2016, Dartois2019}). Hence, CH$_{3}$OH can probe different mechanisms that can inject it into the gas phase, such as thermal desorption, shocks, irradiation by UV photons, and cosmic ray bombardment. In this vein, we present the first study of the filamentary structures and shocks within the protostellar cluster OMC-2\,FIR\,4 at an unprecedented angular resolution of $\sim$100 au, where we targeted eight different molecular tracers, including SiO and CH$_{3}$OH, with the Atacama Large (sub-)Millimetre Array (ALMA) at 1.2 mm.
\newline \indent
The paper is structured as follows: In Sect. \ref{Obs_and_lines} we describe the observations.\ In Sect. \ref{Results} we present the molecular line maps, together with the main results of the analysis. In Sect. \ref{Discussions} we discuss the results, and, finally, in Sect. \ref{Conclusions} we summarise the conclusions.

\section{Observations}

\label{Obs_and_lines}

OMC-2\,FIR\,4 was observed with ALMA during its Cycle 4 operations, between 25 October 2016 and 5 May 2017, as part of the project 2016.1.00376.S (PI: Ana López-Sepulcre). The observations were performed using the ALMA main array in its C-5 configuration, probing angular scales from 0$\farcs$19 to 11$\farcs$2. Several spectral windows (spw) were placed within the spectral range (218-234 GHz and 243-262 GHz). They were observed using spectral channels of 122.070 kHz ($\sim$ 0.14--0.17 km s$\rm ^{-1}$). A subsequent spectral smoothing by a factor 3 was applied for some spws. The phase-tracking centre was R.A. (J2000)$ = 05^{h}35^{m}26^{s}.97$, Dec. (J2000)$ = -05^{\circ}09'54''.50,$ and the systemic velocity was set to $V_{LSR}$ = 11.4 km s$^{-1}$. The quasars J0510+1800 and J0522-3627 were used for bandpass and flux calibration, while J0607-0834 and J0501-0159 were used for phase calibration. The absolute flux calibration uncertainty is estimated to be $<$10\%. \newline The data calibration was performed using the standard ALMA calibration pipeline with the Common Astronomy Software Applications package (CASA\footnote{\url{https://casa.nrao.edu/}}; \citealp{McMullin2007}), while self-calibration, imaging, cleaning, and data analysis were performed using the IRAM-GILDAS software package\footnote{http://www.iram.fr/IRAMFR/GILDAS/}. The continuum images were produced by averaging line-free channels from the wide spectral windows at 232 GHz and 246 GHz in the visibility plane. 
The resulting beam size and rms are (0$\farcs$28 $\times$ 0$\farcs$23 and 95.7 $\mu$Jy\,beam$^{-1}$) and (0$\farcs$50 $\times$ 0$\farcs$27 and 91.7 $\mu$Jy\,beam$^{-1}$). Phase self-calibration was performed on the continuum emission, and the gain solutions were applied to the line cubes. Continuum subtraction was performed on the cubes in the visibility plane before line imaging. The data were cleaned in GILDAS MAPPING using the Clark algorithm \citep{Clark1980}. The resulting synthesised clean beam, velocity resolution, and channel rms for each spectral window are summarised in Table \ref{spw}. The line identification was performed using the Cologne Database for Molecular Spectroscopy (CDMS\footnote{\url{https://cdms.astro.uni-koeln.de}}; \citealt{Muller2001,Muller2005}). The maps shown in the paper are not corrected for the primary beam attenuation because we are more interested in the structure morphology (rather than the flux) that was not affected by the primary beam attenuation. However, we took the correction into account in the spectra.


%
{\footnotesize\setlength{\tabcolsep}{3pt}
\begin{table*}[t]
  \caption{List of transitions, their parameters, and spectral window parameters.}
  \renewcommand{\arraystretch}{1.2}
  \begin{tabular}{l l c c c c c c c }
  
    \hline
        \hline
         
         $\mathrm{Molecule}$ & 
         $\mathrm{Transition}$ &
         $\mathrm{\nu \, ^{a}} $ &
         $\mathrm{E_{up} \, ^{a}}$ &
         $\mathrm{S_{ij} \, \mu ^{2} \, ^{a}}$ & $n_{cr} \, ^{c}$&
        $\mathrm{dV} $ &
         $\mathrm{Beam \, (PA)}$ &
         $\mathrm{Chan. \, rms} $  \\  
         
               & &
         (MHz) &
         (K) & (D$^{2}$)& (10$^{4}$ cm$^{-3}$) &
          (km s$\rm ^{-1}$) & ($\arcsec$ $\times$ $\arcsec$, $\rm ^{\circ}$) &
         (mJy beam$\rm ^{-1}$) \\         \hline

HC$_{3}$N & 24-23 & 218324.72 & 131 & 334.2 & 8.3& 0.5 $^{d}$ & 0.52 $\times$ 0.29 ($-$69) & 5.4 \\

CH$_{3}$OH & 4$_{-2,3}$ $-$ 3$_{-1,2}$ \, E & 218440.06 & 45 & 13.9 & 2.2 & 0.17 & 0.55 $\times$ 0.29 ($-$73) & 5.5 \\ 
 & 4$_{2,3}$ $-$ 5$_{1,4}$ \, A & 234683.37 & 61 & 4.5 & 0.5 & 0.16 & 0.49 $\times$ 0.27 ($-$71) & 5.8 \\ 
 & 5$_{1,4}$ $-$ 4$_{1,3}$ \, A & 243915.78 & 50 & 15.5 & 1.9 & 0.15 & 0.27 $\times$ 0.22 ($-$95) & 4.7 \\

C$^{18}$O & 2-1 & 219560.35 & 16 & 0.02 & 0.5 & 0.17 & 0.53 $\times$ 0.29 ($-$71) & 5.0 \\

CS & 5-4 & 244935.56 & 35 & 19.1 & 130  & 0.15 & 0.33 $\times$ 0.27 ($-$68) & 3.7 \\

$\mathrm{CH_{3}CN}$ & 14$\rm{_{1}}-$ 13$\rm{_{1}}$ & 257522.42 & 100 & 295.2 & 9.3 & \multirow{2}{*}{ 0.42 $^{d}$} & \multirow{2}{*}{0.31 $\times$ 0.26 (-71)}& \multirow{2}{*}{2.8} \\ 
 & 14$\rm{_{0}}-$ 13$\rm{_{0}}$ & 257527.38 & 93 & 296.7 & 10.2  \\

H$^{13}$CN \, $\mathrm{^{b}}$ & J=3$-$2, F=4$-$3 & 259011.86 & 25 & 34.4 & 410 & 0.45 $^{d}$ & 0.33 $\times$ 0.27 ($-$68) & 2.1  \\

SiO & 6-5 & 260518.02 & 44 & 57.6 & 50 & 0.42 $^{d}$ & 0.30 $\times$ 0.26 ($-73$) & 2.5 \\
  
CCH \, $\mathrm{^{b}}$ & N= 3$-$2, J=5$/$2$-$3$/$2, F=3$-$2 & 262064.99 & 25 & 4.3 & 6.5 &  0.42 $^{d}$ & 0.30 $\times$ 0.26 ($-$73) & 2.5 \\

  \hline
  \end{tabular}
  \label{spw}
  \noindent
\textit{\textbf{Notes:}} \newline $^{a}$ Frequencies and spectroscopic parameters have been extracted from the CDMS catalogue \citep{Muller2001,Muller2005}.  \newline $^{b}$ H$^{13}$CN and CCH consist of two hyperfine components in 0.4 MHz interval. Those with the higher S$_{ij} \, \mu^{2}$ are reported here. \newline $^{c}$ The critical densities of CS and H$^{13}$CN were obtained from \cite{Shirley2015} at 50 K, and those of the other tracers were estimated at 50 K using the LAMDA database \citep{Schoier2005}. \newline $^{d}$ The spectral channels of these spws were smoothed by a factor of 3 to gain in signal-to-noise ratio.
\end{table*}}


\section{Results}

\label{Results}

We have imaged the emission from eight different molecules: SiO, CH$_{3}$OH, C$^{18}$O, CS, CH$_{3}$CN, H$^{13}$CN, HC$_{3}$N, and CCH at a high angular and spectral resolution ($\sim$0$\farcs$33 and $\sim$0.16 km\,s$^{-1}$). We mapped three different transitions of CH$_{3}$OH and one transition for the other molecules (see Table \ref{spw}). 

\subsection{Spatial distribution of molecular emission}
\label{morphology}

\subsubsection{CS and C$^{18}$O}

In Fig. \ref{mom0-all} we show the velocity-integrated maps\footnote{The maps were integrated between the following ranges: 7.4 and 15.4 km\,s$^{-1}$ for C$^{18}$O, -0.2 and 24.4 km\,s$^{-1}$ for CS, -6.2 and 24.4 km\,s$^{-1}$ for SiO, 3.4 and 19.4 km\,s$^{-1}$ for CH$_{3}$OH, 7.6 and 23.6 km\,s$^{-1}$ for CH$_{3}$CN, 1.5 and 21.8  km\,s$^{-1}$ for H$^{13}$CN, 5 and 21.4  km\,s$^{-1}$ for HC$_{3}$N, and 2.1 and 20.6 km\,s$^{-1}$ for CCH.} of the observed molecular tracers in colours, on which we overlaid CH$_{3}$OH (5$_{1,4}-$4$_{1,3}$\,A) contours. CS, a typical high-density tracer ($n_{cr}$=1.3$\times$10$^{6}$ cm$^{-3}$ for the 5-4 transition, see Table \ref{spw}), shows the brightest and most extended emission in our maps. It probes the overall structures of the gas in the FIR\,4 protocluster, a ridge in the east side extending from the north-east (NE) to the south-west (SW) (hereafter called eastern ridge), and several filamentary structures. The eastern ridge has a length of $\sim$20$\arcsec$ ($\sim$7900 au) and extends over $\sim$3300 au. The other filamentary structures are relatively shorter, with a length ranging between $\sim$4$\arcsec$ and 10$\arcsec$ ($\sim$1500-4000 au). 
C$^{18}$O traces the eastern ridge, but its emission is less extended (see Sect. \ref{kinematics}). 
All the other molecules are emitting from the region covered by CS, except for SiO. 

\subsubsection{SiO}

SiO traces the shocked regions within the FIR\,4 protocluster. From the overlap of its velocity-integrated map with the CH$_{3}$OH contours, we can notice a different morphology and a shift in the emission. More specifically, two regions of the cloud can be distinguished: a region dominated by SiO located to the west, and another dominated by CH$_{3}$OH to the east. Furthermore, SiO reveals multiple bow-shock features: an arc-like extended bow shock to the west of the cluster (hereafter called Arc\,1) and a smaller bow shock located south-west of the cluster (hereafter called arc\,2). Arc\,1 extends over $\sim$2700 au, while arc\,2 extends over $\sim$500 au. SiO emission had already been reported towards the west of FIR\,4 region, but at lower angular resolution ($\sim$3$\arcsec-$7$\arcsec$; \citealt{Shimajiri2008}). In addition, we report a collimated monopolar molecular jet from the protostar VLA15. The jet extends along the S-N direction, with a length of 13.3$\arcsec$ ($\sim$5200 au). The source driving the jet was recently identified as a young and deeply embedded protostar \citep{Osorio2017, Tobin2019} that appears to have an edge-on horizontal disk \citep{Tobin2019}. The class of this protostar has not yet been determined.


\subsubsection{CH$_{3}$OH}

\begin{figure*}
    \centering
    \includegraphics[width=0.86\textwidth]{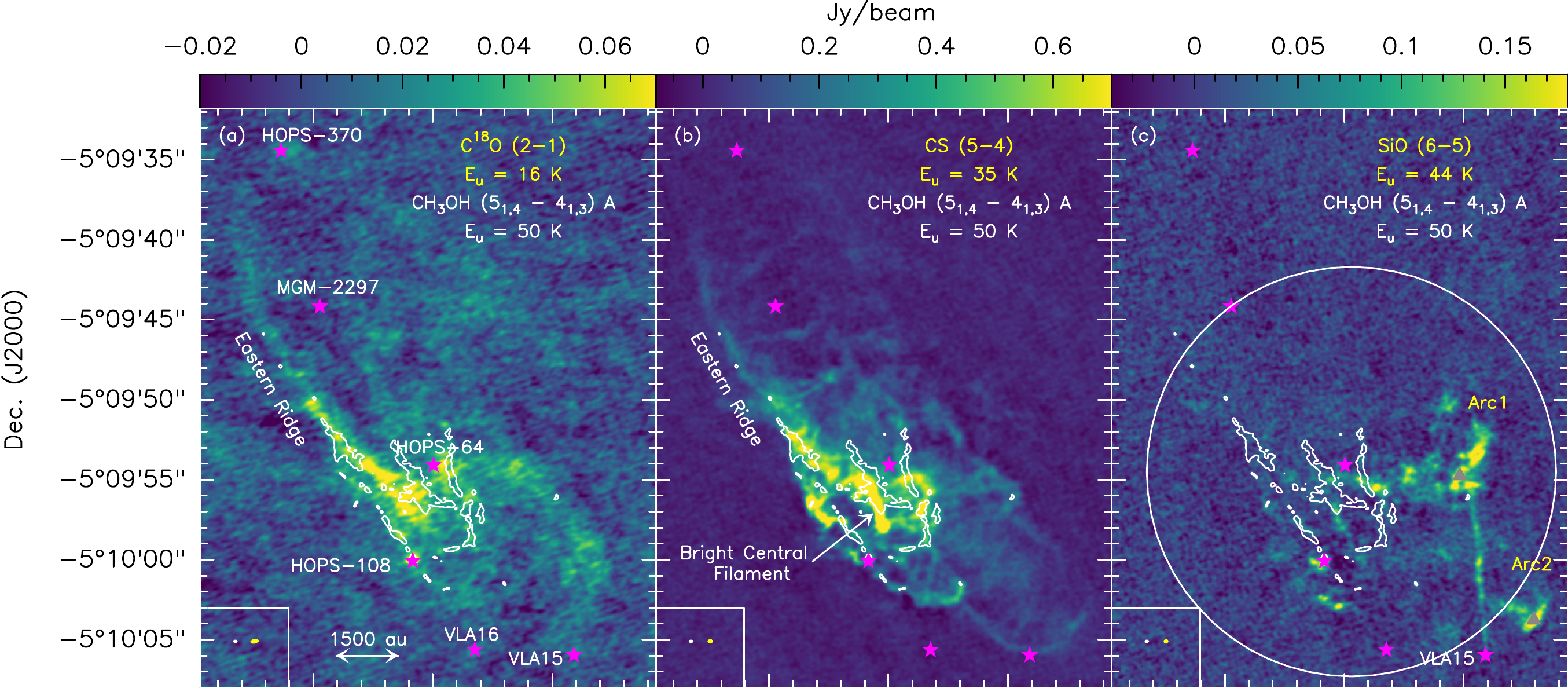}
    \includegraphics[width=0.86\textwidth]{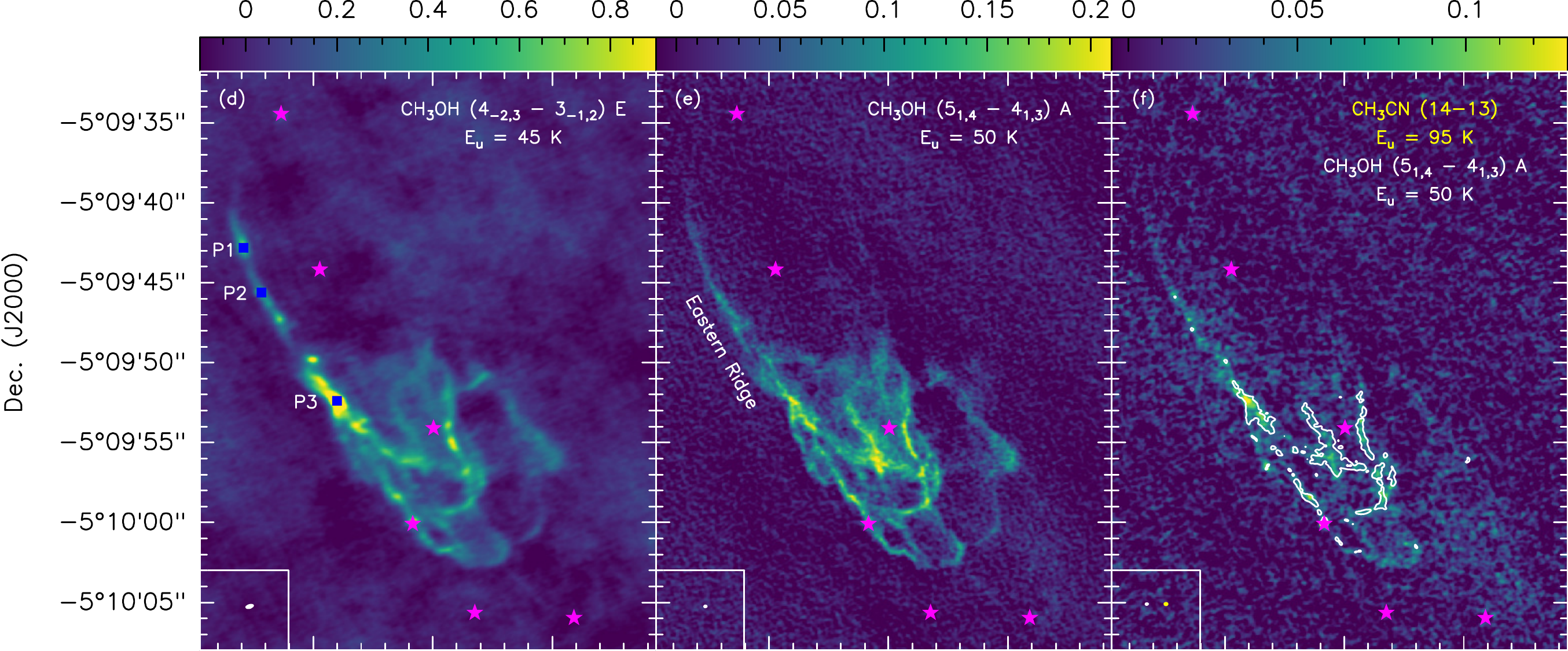}
    \includegraphics[width=0.86\textwidth]{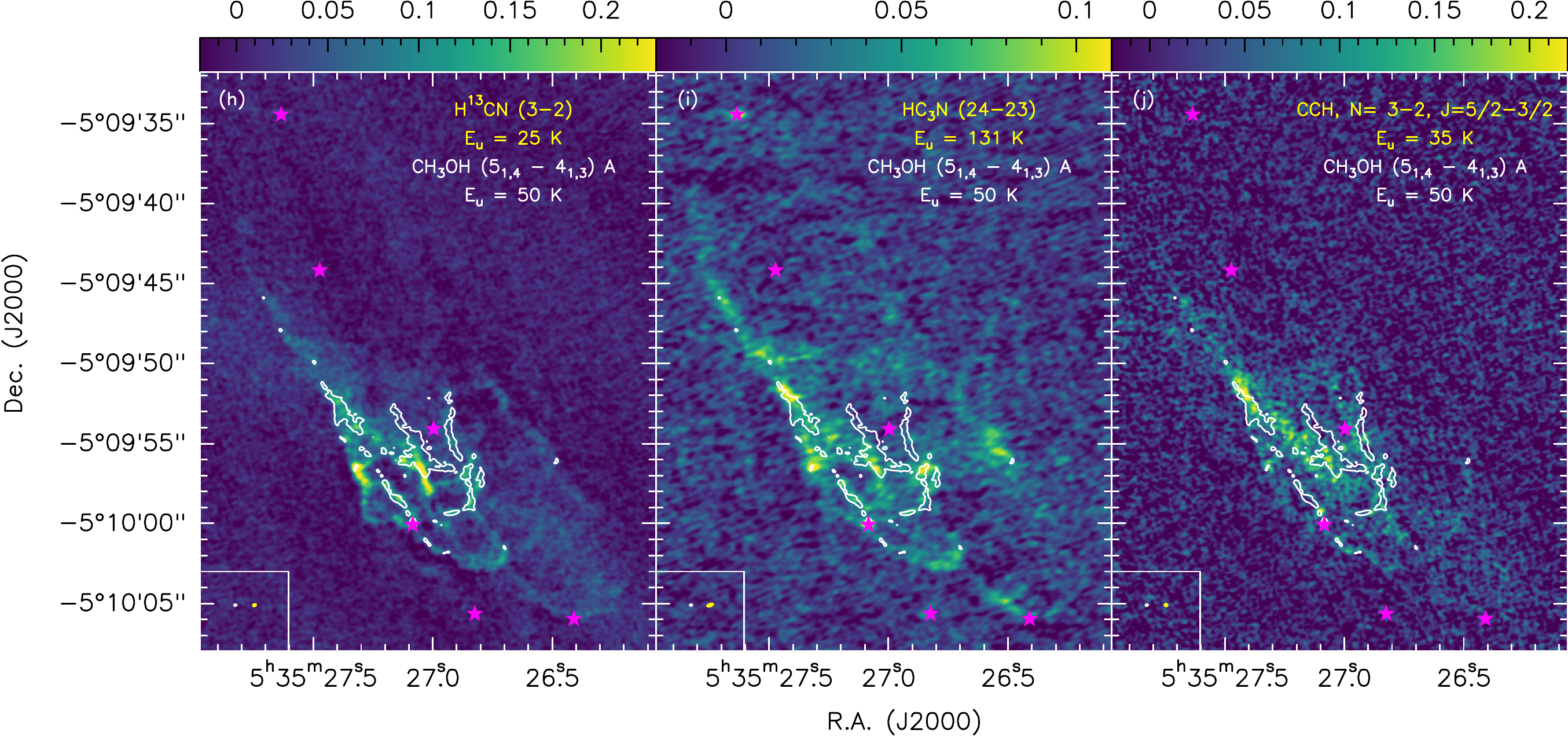}
    \caption{Velocity-integrated maps of the different tracers used to probe the emission towards OMC-2\,FIR\,4.\newline \textit{Upper panels}: Velocity-integrated maps of C$^{18}$O, CS and SiO in colours, with superimposed contours of the velocity-integrated emission of CH$_{3}$OH (5$_{1,4}-$4$_{1,3}$\,A) line (white). The colour images are shown at the top of each panel.
    The different cores are depicted with magenta stars and are labelled in white. The synthesised beam of CH$_{3}$OH and the other molecules are depicted in the lower left corner in white and yellow, respectively. In the SiO panel, the positions of the SiO peaks from this study are depicted with filled grey triangles. The primary beam at the full width at half maximum is depicted in white. \newline \textit{Middle panels}: Velocity-integrated maps of CH$_{3}$OH lines at E$_{u}$=45 K and E$_{u}$=50 K and of CH$_{3}$CN. On the latter, we superimposed the contours of the velocity-integrated emission of CH$_{3}$OH (5$_{1,4}-$4$_{1,3}$\,A) line (white). The colour images are shown at the top of each panel.
    The positions of the three masers detected in CH$_{3}$OH (4$_{-2,3}-$3$_{-1,2}$\,E) are depicted with filled blue squares. \newline \textit{Lower panels}: Velocity-integrated maps of H$^{13}$CN, HC$_{3}$N and CCH in colours with superimposed contours of the velocity-integrated emission of CH$_{3}$OH (5$_{1,4}-$4$_{1,3}$\,A) line (white). The colour images are shown at the top of each panel
    The methanol contours start at 10$\sigma$ with $\sigma$ = 12.3 mJy beam$^{-1}$ km\,s$^{-1}$.}
   \label{mom0-all}
\end{figure*}

Similarly to CS, the CH$_{3}$OH velocity-integrated maps shown in the middle panel of Fig. \ref{mom0-all} reveal several cavities from NE to SW and widespread interwoven filamentary structures. The origin of these structures is discussed in Sect. \ref{fil_orgin}. 

For the CH$_{3}$OH (4$_{-2,3}-$3$_{-1,2}$\,E) transition, known to be a class I-type CH$_{3}$OH maser \citep{Hunter2014,Chen2019}, we noted that in three positions (P1, P2, and P3) along the eastern ridge (see panel (d) in Fig. \ref{mom0-all}), the line intensities were >300 K km\,s$^{-1}$. This suggests the presence of maser spots at these positions. Maser spots were previously identified at the same position, with different CH$_{3}$OH maser transitions (e.g. \citealt{Kogan1998, Kang2013}; Fontani et al. in prep). The maser at P2 was observed at 7 mm, 3 mm, and 2 mm \citep{Kang2013}, while those at P1 and P3 were observed at 3 mm (Fontani et al. in prep). Hence, we report the detection of these masers at 1 mm. The locations of the maser spots we detect along the eastern ridge are reported in Table. \ref{maser-pos}. The narrow spectra extracted at the peak of the above-mentioned spots are shown in Fig. \ref{masers-spec} of Appendix \ref{spectra}.
It is worth noting that on one hand, the class I-type CH$_{3}$OH masers result from collisional pumping (e.g. \citealt{Sobolev2007, Ladeyschikov2020, Nesterenok2022}), which can be associated with young stellar object outflows \citep{Voronkov2006, Chen2011, Ladeyschikov2020}, expanding H\,II regions \citep{Voronkov2014}, expanding supernova remnants \citep{Pihlstrom2014}, and rapidly moving cloudlets \citep{Voronkov2010}. Hence, they indicate the presence of compressed gas. On the other hand, \cite{Kang2013} suggested that the exciting source of the maser at P2 could be one of the protostars in the OMC-2 FIR\,3/4 clusters, and \cite{Osorio2017} have reported radio emission associated with the jet from the protostar HOPS-370 close to the maser at P3. 

{\footnotesize\setlength{\tabcolsep}{12pt}
\begin{table}
  \caption{Position of the detected CH$_{3}$OH (4$_{-2,3}-$3$_{-1,2}$\,E) maser spots along the eastern ridge.}
   \renewcommand{\arraystretch}{1.2}
  \begin{tabular}{c c c}
    \hline \hline
        
         $\mathrm{Spot}$ & 
         $\mathrm{R.A.(J2000)}$ &
         $\mathrm{Dec.(J2000)} $ \\          
         
         & ($^{h \; m \;s}$) & ($^{\circ}$ ${\: '\: ''}$) \\
         
         \hline
         
        P1 & 05:35:27.792 & $-$05:09:42.82 \\
    
        P2 & 05:35:27.717 & $-$05:09:45.62 \\
        
        P3 & 05:35:27.401 & $-$05:09:52.38 \\ \hline
        
  \end{tabular}
  \label{maser-pos}
  \noindent
\end{table}}

We detected three CH$_{3}$OH lines with close upper state energies. Therefore, we were not able to perform a non-local thermodynamic equilibrium analysis (non-LTE). 

\subsubsection{H$^{13}$CN, CH$_{3}$CN, HC$_{3}$N, CCH, and dust}


The emission of H$^{13}$CN, CH$_{3}$CN, HC$_{3}$N, and CCH is fainter, but agrees reasonably well with the CS and CH$_{3}$OH morphology. The eastern ridge is detected in all the tracers. The hot corino HOPS-108 \citep{Tobin2019, Chahine2022} is also observed in all the molecules except for CCH, which is a traditional probe of UV illuminated regions (e.g. \citealt{Gratier2017, Bouvier2020}). Moreover, the bright central filament ($\sim$4$\arcsec$ in length) probed by CS is very well detected in H$^{13}$CN, the other tracer of high-density gas ($n_{cr}$=4.1$\times$10$^{6}$ cm$^{-3}$ for the 3-2 transition, see Table \ref{spw}), suggesting that this filament might be one of the densest within the protocluster.

\begin{figure}
    \centering
    \includegraphics[width=0.48\textwidth]{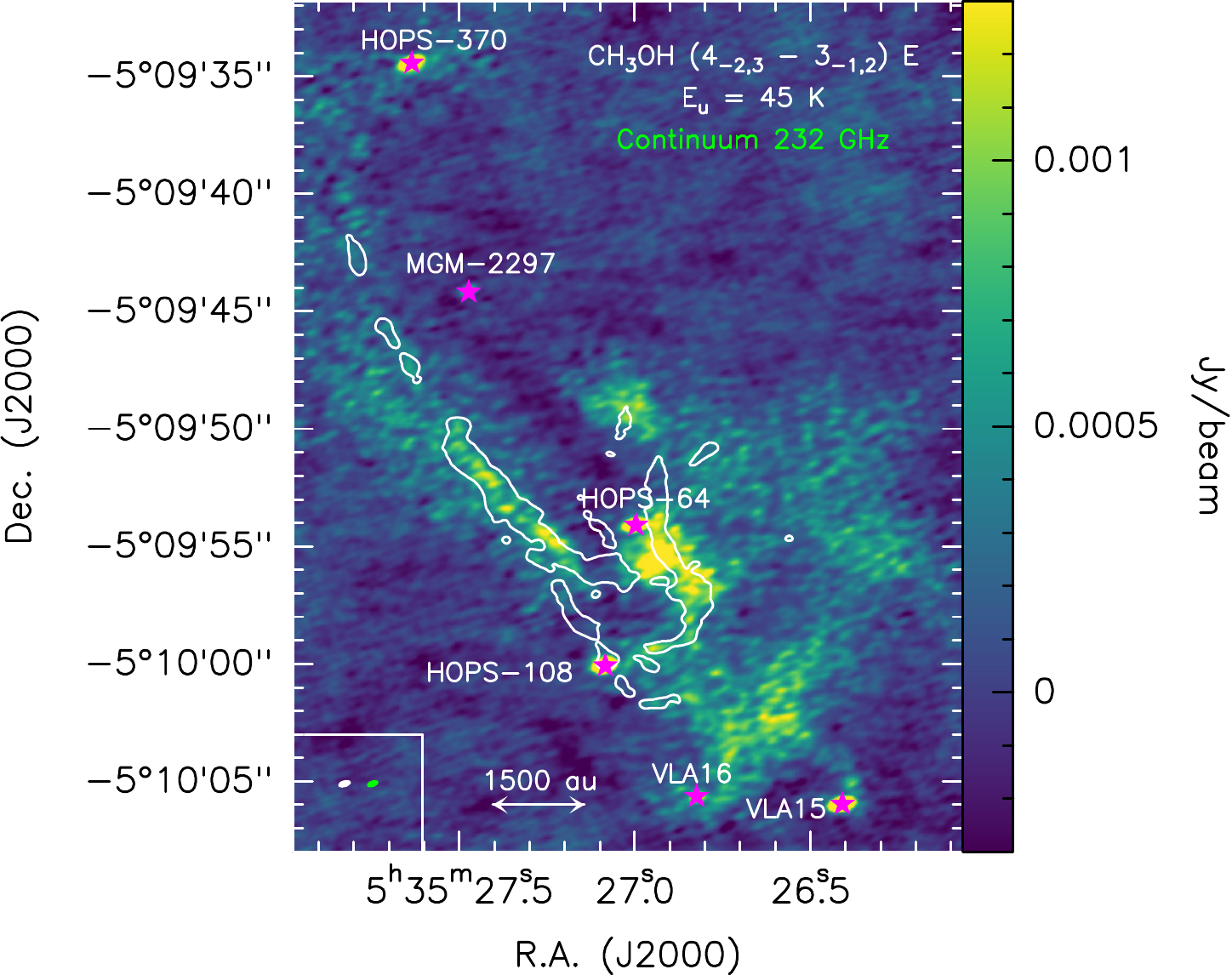}
    \caption{Continuum emission towards OMC-2 FIR\,4 at 232 GHz in colours, with superimposed contours of the velocity-integrated emission of CH$_{3}$OH (4$_{-2,3}-$3$_{-1,2}$\,E) at 15$\sigma$ with $\sigma$ = 20.6 mJy\,beam$^{-1}$\,km\,s$^{-1}$. The different cores are depicted with magenta stars and are labelled in white. The synthesised beam of CH$_{3}$OH and continuum are depicted in the lower left corner in white and green, respectively.}
    \label{cont-meth}
\end{figure}

Finally, in addition to probing the protostars within OMC-2\,FIR\,4, the continuum emission also traces some filamentary structures. The overall distribution of CH$_{3}$OH emission follows the filament morphology observed in the continuum (see Fig. \ref{cont-meth}). The eastern ridge is well probed by dust emission, suggesting that it is in high-density conditions.

\subsection{Kinematics of molecular emission}
\label{kinematics}

\begin{figure*}
   
        \includegraphics[width=1\textwidth]{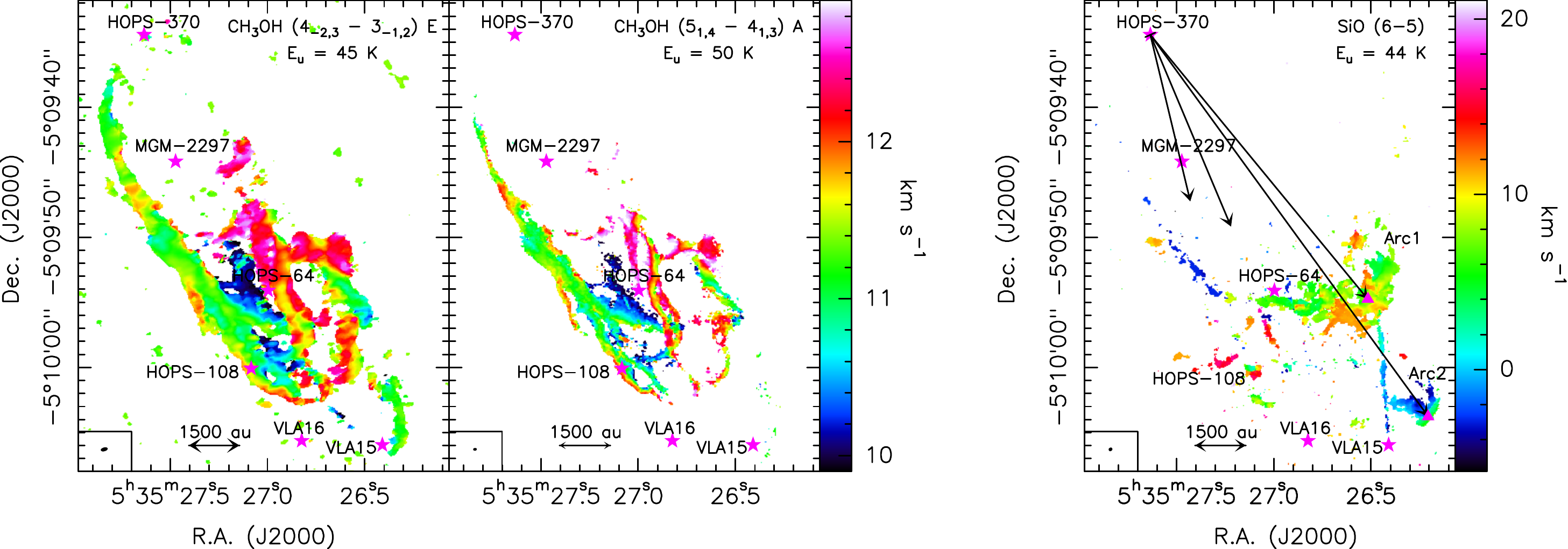} %

\caption{Velocity maps of CH$_{3}$OH and SiO towards OMC-2\,FIR\,4. \newline Left panel: Velocity maps (integrated between 9.9 and 12.9 km\,s$^{-1}$; colour scale in km\,s$^{-1}$) of CH$_{3}$OH (4$_{-2,3}-$3$_{-1,2}$\,E) (left) and CH$_{3}$OH (5$_{1,4}-$4$_{1,3}$\,A) (right). Right panel: Velocity map (integrated between -6.2 and 24.4 km\,s$^{-1}$; colour scale in km\,s$^{-1}$) of SiO(6-5). The different cores are depicted with magenta stars and are labelled in black. The positions of the SiO peaks from this study are depicted with filled magenta triangles. We draw four arrows from HOPS-370, one perpendicular to the disk, two towards the SiO peaks, and one similar to the direction of the jet in [O\,I] by \cite{Gonzalez-Garcia2016}.}
\label{moms-1-all}    
\end{figure*}

We provide velocity maps (moment 1) and velocity-dispersion maps (moment 2) for CH$_{3}$OH and SiO (Figs. \ref{moms-1-all} and \ref{moms-2-all}). We also show the velocity channel maps for CH$_{3}$OH, C$^{18}$O, HC$_{3}$N, CS, and SiO (see Appendix \ref{app}, Figs. \ref{comp-red-blu-meth}-\ref{sio-chan-map}).
C$^{18}$O and CS are plausibly affected by missing flux due to structures larger than the maximum recoverable scale (MRS$\sim$11$\farcs$2). Therefore, we did not analyse the kinematics using these tracers. The remaining molecular tracers are either faint or have a hyperfine structure, hence the study of their kinematics would not be reliable.

The average systemic velocity of the FIR\,4 protocluster is V$_{sys}$=11.4 km\,s$^{-1}$ \citep{Shimajiri2008, Favre2018}. Remarkably, from the velocity maps and channel maps of CH$_{3}$OH, we can clearly see that each filament has a different velocity and that the filaments at the centre of the cloud have higher velocities ($>$2 km\,s$^{-1}$) than the external filaments in both red- and blueshifted emission. The velocity gradient in CH$_{3}$OH between the east and the west of the FIR\,4 protocluster is about $\pm$4 km\,s$^{-1}$ with respect to the systemic velocity. 
From the velocity-dispersion maps of CH$_{3}$OH in Fig \ref{moms-2-all}, it seems that we have three distinct spatial distributions of the velocity dispersion: 
\begin{enumerate}[nolistsep]
    
    \item The external filamentary structures that have the lowest velocity dispersion, $<$2 km\,s$^{-1}$.
    \item The intermediate filaments around the central region that have a velocity dispersion of 2-4 km\,s$^{-1}$.
    \item The central region showing (in the moment 2 map) a velocity dispersion of 4-7 km\,s$^{-1}$.
\end{enumerate}
However, the high dispersion in the central region is actually due to the presence of multiple structures along the line of sight (see Appendix \ref{spectra}, Fig. \ref{meth-spec}). In Fig. \ref{meth-spec} we show the spectra extracted from three different positions that are marked by magenta triangles in the moment-2 map.

 \vspace{2.5mm}
A completely different velocity distribution is observed in SiO. Its emission extends up to much higher blueshifted velocities than the other tracers, covering a wide velocity range of $-$17.6 $<$ V-V$_{sys}$ $<$ +8 km\,s$^{-1}$. From the SiO velocity and channel maps, three velocity regimes with different spatial distribution can be identified: 
\begin{enumerate}[nolistsep]
    \item Blueshifted high-velocity (HV) regime in the range $-$17.6 $<$ V-V$_{sys}$ $<-$8.1 km\,s$^{-1}$, where SiO reveals the decelerating jet from VLA15, the south-west bow-shock (arc\,2) and the elongated eastern ridge.
    \item Blueshifted low-velocity (LV) regime in the range $-$8 $<$ V-V$_{sys}$ $<$ 0 km\,s$^{-1}$, where SiO probes the extended arc-like bow (arc\,1) west of the cluster.    \item Redshifted low-velocity (LV) regime in the range +0.1 $<$ V-V$_{sys}$ $<$ +8 km\,s$^{-1}$, where SiO probes more compact filaments than the blueshifted parts.
\end{enumerate}

\vspace{2mm}
It is worth noting that the redshifted counterpart of the jet from VLA15 is not detected in our data as it is $\sim$3$\arcsec$ out of our primary beam. However, the possibility of having a monopolar jet cannot be excluded (e.g. \citealt{Codella2014}). Interestingly, the jet can also be observed in CS at high blueshifted velocities in the range $-$0.2$<$ V$<+$3.3 km\,s$^{-1}$ (see the bottom left panel in Fig. \ref{comp-red-blu-cs} of  Appendix \ref{app}).

We overlaid the integrated contours of each velocity regime on the velocity-integrated maps of CH$_{3}$OH, CS, and H$^{13}$CN (see Fig. \ref{sio-others}). We note that the bow shocks in SiO are spatially unrelated to the bulk emission that is traced by the other molecules, but SiO traces some filamentary structures within OMC-2\,FIR\,4 at certain velocities. At higher blueshifted velocities (HV) SiO traces the eastern ridge, which is probed by all the other molecular tracers and by the dust emission. It also traces the filamentary structure located west of the protostar HOPS-64 and the cavity located west of the source HOPS-108. While in the red LV regime, SiO probes the central collimated filament, which is also observed in CS and H$^{13}$CN. The detection of SiO in some filamentary structures indicates that these filaments contain shocked gas, where the gas is compressed and warmed up. This is also confirmed by the detection of the class-I CH$_{3}$OH maser spots along the eastern ridge.

\begin{figure*}
    \centering
        \includegraphics[width=0.92\textwidth]{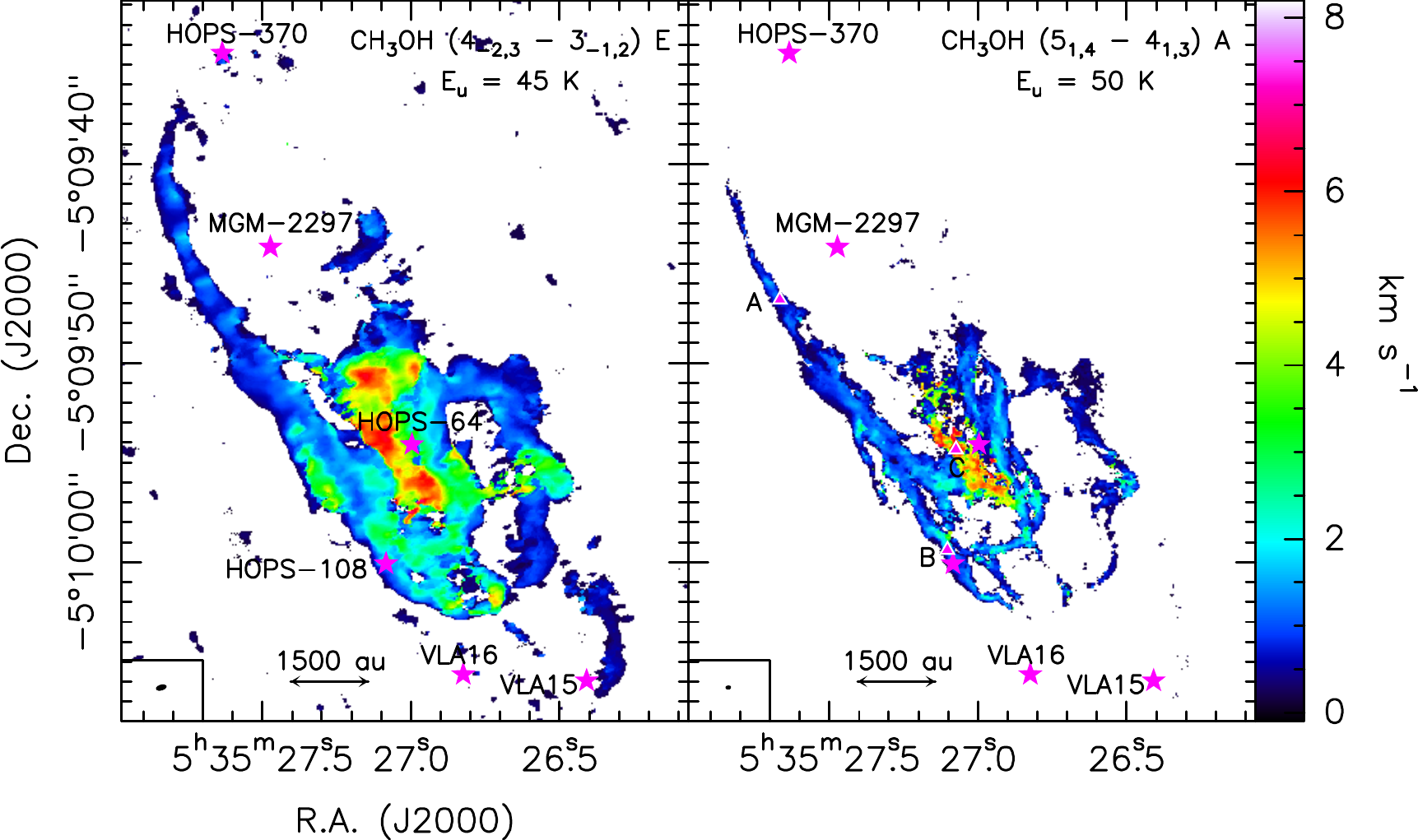}

\caption{Velocity-dispersion maps (integrated between 7.4 and 15.4 km\,s$^{-1}$; colour scale in km\,s$^{-1}$) of CH$_{3}$OH (4$_{-2,3}-$3$_{-1,2}$\,E) (left) and CH$_{3}$OH (5$_{1,4}-$4$_{1,3}$\,A) (right). The positions (A, B, and C) from which we extracted the spectra in Fig. \ref{meth-spec} are labelled with filled magenta triangles.}
\label{moms-2-all}    
\end{figure*}

\begin{figure*}
    \centering
    \includegraphics[width=1\textwidth]{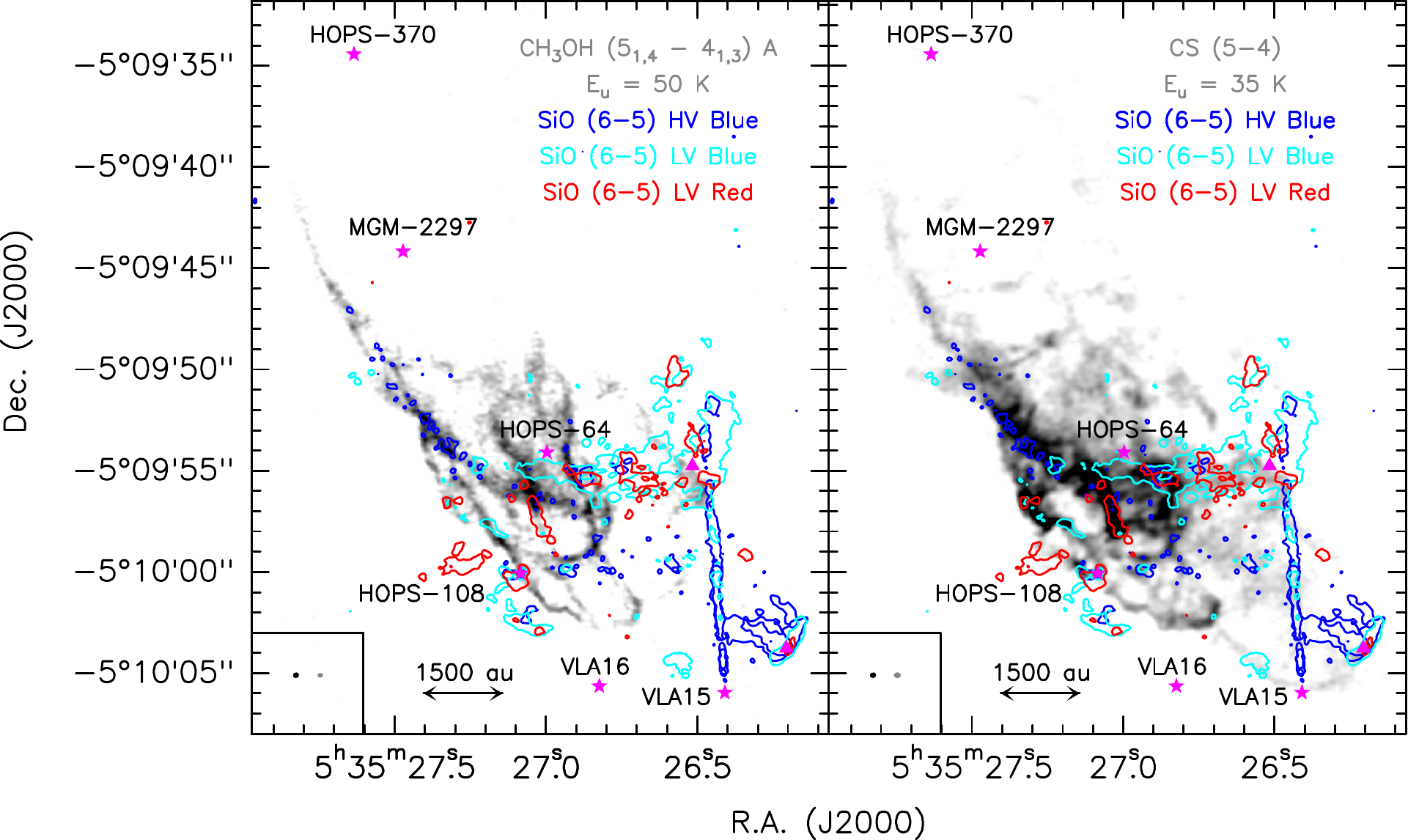}
    \caption{Velocity-integrated maps of CH$_{3}$OH and CS reported in Fig. \ref{mom0-all} with superimposed contours of the velocity-integrated emission of SiO at different velocity regimes. The colour images are as in Fig. \ref{mom0-all}. The SiO contours are at 5$\sigma$ with $\sigma$= 5.8 mJy\,beam$^{-1}$\,km\,s$^{-1}$. They correspond to the blueshifted high-velocity regime (blue), the blueshifted low-velocity regime (white), and the redshifted low-velocity regime (red) (see Sect. \ref{kinematics}). The different cores are depicted with magenta stars and are labelled in black. The positions of the SiO peaks from this study are depicted with filled magenta triangles.}
   \label{sio-others}
\end{figure*}

\begin{figure*}
    \centering
     \includegraphics[width=0.48\textwidth]{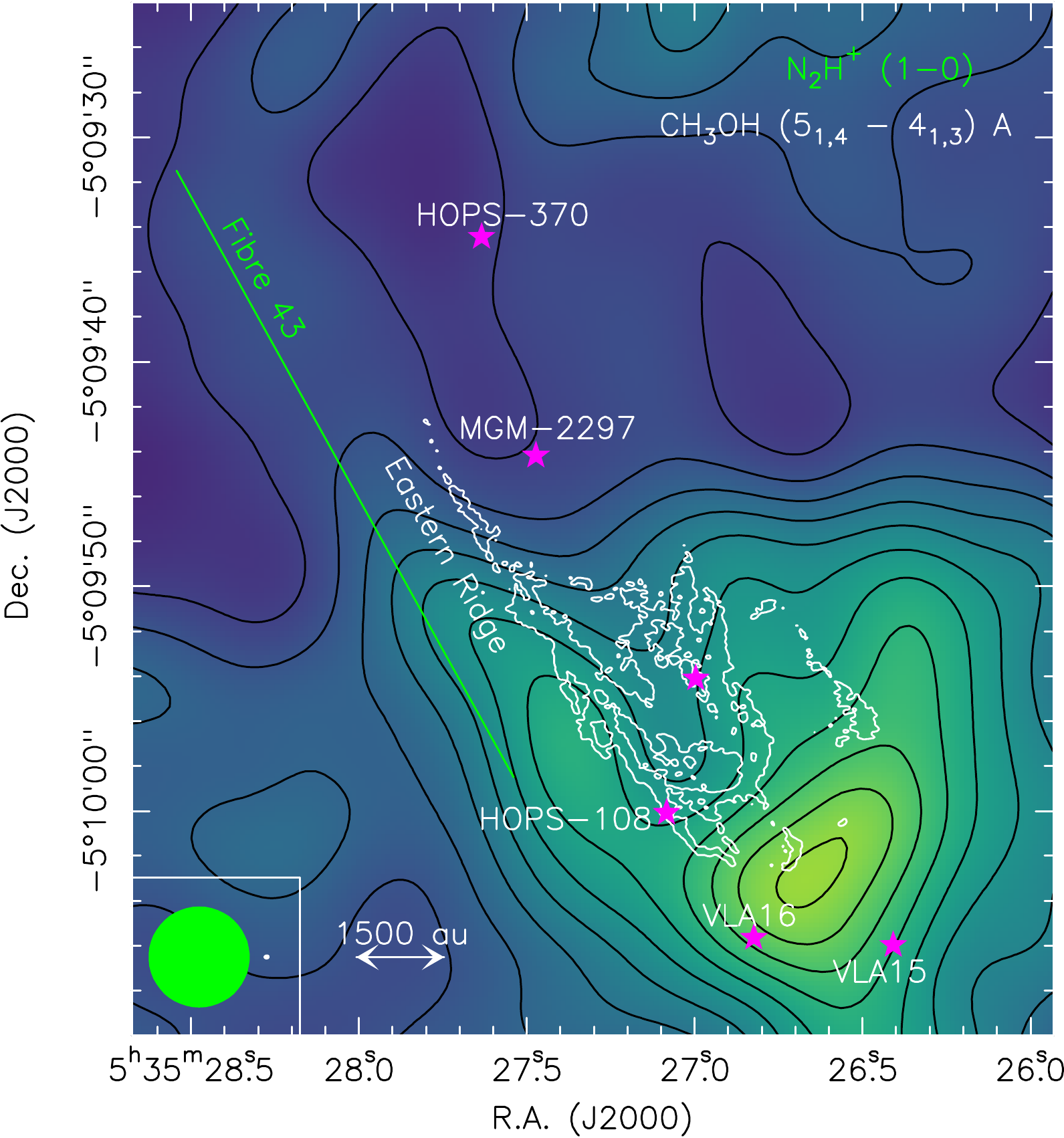}
     \includegraphics[width=0.48\textwidth]{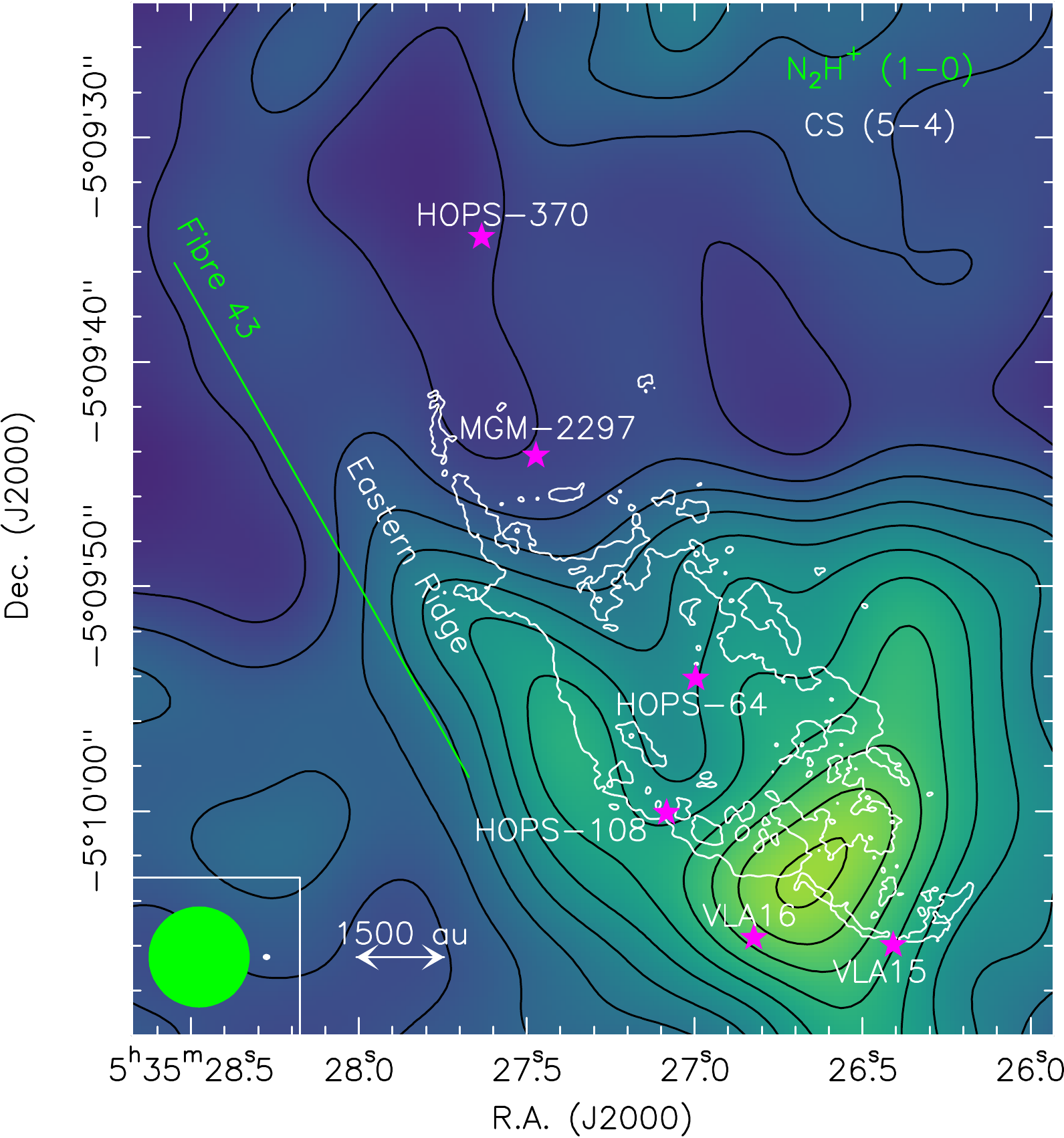}
      \caption{N$_{2}$H$^{+}$(1--0) integrated intensity emission from \cite{Hacar2018} in colour scale and black contours, with superimposed 5$\sigma$ contours of the velocity-integrated emission of CH$_{3}$OH (5$_{1,4}-$4$_{1,3}$\,A) (left) and CS (right) from this work. N$_{2}$H$^{+}$ contours are equally spaced every 1 Jy\,beam$^{-1}$\,km\,s$^{-1}$. The CH$_{3}$OH and CS contours are shown at 5$\sigma$ with $\sigma$= 12.3 mJy\,beam$^{-1}$\,km\,s$^{-1}$ and 23.2 mJy\,beam$^{-1}$\,km\,s$^{-1}$, respectively. The different cores are depicted with magenta stars and are labelled in white. The direction of Fibre 43 is shown with a green line. The synthesised beams of N$_{2}$H$^{+}$, CH$_{3}$OH (left), and CS (right) are depicted in the lower left corner in green and white, respectively.}
      \label{large-small-scale}
\end{figure*}


\section{Discussion}

\label{Discussions}


\subsection{Origin of the filaments}
\label{fil_orgin}

At $\sim$1 pc scale and $\sim$1800 au angular resolution, \cite{Hacar2018} identified 55 fibres along the OMC-1 and OMC-2 clouds of the ISF using the quiescent gas tracer N$_{2}$H$^{+}$. They have suggested that the fibres towards the OMC-1 south region, the OMC-1 ridge, and towards the centres of OMC-2 FIR\,4 and OMC-2 FIR\,6 are distributed in fan-like arrangements, while they are sparse outside these regions (see their Sect. 3.5). Furthermore, they noted that the fibre hubs coincide with the positions of different (stars plus gas) mass concentrations along the ISF. The algorithm HiFive used in their study was not able to decompose the most complex structures, such as OMC-2 FIR\,4, which consists of multiple sub-branches. However, one main fibre within this region was clearly identified in their study, and it is fibre number 43 (see their Fig. 4). 

Our observations of the OMC-2 FIR\,4 protocluster at a spatial resolution of $\sim$100 au reveal a collection of intertwined filamentary structures within a field of view of $\sim$10000 au. To assess whether these structures are associated with the fibres observed with N$_{2}$H$^{+}$ at larger scales, we superimposed the CH$_{3}$OH (5$_{1,4}-$4$_{1,3}$\,A) and CS contours emission from our work on the  N$_{2}$H$^{+}$ (1--0) emission map from \cite{Hacar2018} (see Fig. \ref{large-small-scale}). This figure shows that N$_{2}$H$^{+}$ emission surrounds that of CH$_{3}$OH and CS. Our eastern ridge is aligned with fibre 43 identified by \cite{Hacar2018}. The other shorter filaments could be connected to the multiple sub-branches of the fibres' hub that were not recovered by HiFive. Taking into account the differences in scales, angular resolution, and tracers, we can say that the two works are complementary. 
Whereas at 4.5$\arcsec$, \cite{Hacar2018} observed the dense fibres, the higher angular resolution have allowed us to resolve the filamentary structures within the hub. In addition, CH$_{3}$OH has offered us a probe of the dynamics. While with N$_{2}$H$^{+}$ the centroid velocities V$_{LSR}$(N$_{2}$H$^{+}$) of the fitted components in OMC-2 FIR\,4 region range from $\sim$ 10 to 12.5 km\,s$^{-1}$ from east to west with a velocity dispersion between 0.2 and 2.2 km\,s$^{-1}$ (see their Fig. B.3), our CH$_{3}$OH filamentary structures show a higher velocity dispersion, ranging between 0.9 and 3.2 km\,s$^{-1}$ (excluding the line width of maser spots). Each of the CH$_{3}$OH filaments has a different velocity, with those at the centre being higher (V$\sim$ 7.4 km\,s$^{-1}$ for the most blueshifted filament and V$\sim$ 17.4 km\,s$^{-1}$ for the most redshifted one). The velocity range and dispersion are indicative of supersonic motions within the protocluster. To verify this point, we compared the magnitude of the line-of-sight non-thermal velocity dispersion $\sigma_{\rm{NT}}$ with the local thermal sound speed of H$_{2}$, $c_{s}(T_{K})$ (e.g. \citealt{Hacar2011, Hacar2018}). We used the ratio $\sigma_{\rm{NT}}/c_{s}(T_{K})$ to identify whether the observed gas motions are subsonic ($\sigma_{\rm{NT}}/c_{s}(T_{K}) \leq$1), transonic (1$ < \sigma_{\rm{NT}}/c_{s}(T_{K}) \leq$2), or supersonic ($\sigma_{\rm{NT}}/c_{s}(T_{K}) >$2) \citep{Hacar2018,Hacar2022}. The line width of a spectrum combines contributions from both thermal and non-thermal gas motions. Consequently, to estimate the non-thermal component $\sigma_{\rm{NT}}$, we subtracted the thermal component from the measured line width of the spectra, assuming that the two contributions are independent of one another \citep{Myers1983},

\begin{equation}
    \sigma_{\rm{NT}} = \sqrt{ \frac{\Delta V ^{2}}{8 \, \rm {ln} \, 2} - \frac {k_{B} T_{K}}{\mu _{\rm{(CH_{3}OH)}}\,m_{p}}}
.\end{equation}

Thus, for line widths of [0.9--3.2 km\,s$^{-1}$] as obtained from our spectra, $\sigma_{\rm{NT}}$ will range between $\sim$0.4 and 1.4 km\,s$^{-1}$. 
On the other hand, for an isothermal medium, $c_{s}(T_{K})$ is given as
\begin{equation}
    c_{s}(T_{K}) = \sqrt{\frac {k_{B} T_{K}}{\mu _{\rm{(H_{2})}}\,m_{p}}}
.\end{equation}

The kinetic temperature of the gas within OMC-2\,FIR\,4 was estimated to be $\sim$ 40 K \citep{Favre2018}, so $c_{s}(T_{K})$ is $ \sim $ 0.4 km\,s$^{-1}$. Hence, the ratio $\sigma_{\rm{NT}}/c_{s}(T_{K})$ of the filamentary structures ranges between $\sim$ 1 and 3.3, implying the presence of transonic and supersonic motions.
As proposed by \cite{Henshaw2014} for the infrared dark cloud (IRDC) G035.39--00.33, we suggest that the kinematics of the filamentary structures could be affected by the outflows from the forming stars within the protocluster. Gaseous CH$_{3}$OH and CS would testify to a recent injection or formation of these species into the gas-phase, which have not yet had the time to accrete onto the grain mantles. For a density of about 10$^{5}$ cm$^{-3}$ , the timescale for methanol accretion is about 10$^{4}$ yr, and this timescale is slightly longer for CS. This supports the idea that what we see within OMC-2 FIR\,4 could be the result of the protostellar feedback (i.e. outflows, winds, and radiation) that shaped the material in filament-like structures. Hence, the results highlight the importance of stellar feedback on the evolution of molecular hubs, where the material can be shaped and compressed within narrow filaments.

\subsection{SiO fingerprint in filaments}

Another key molecule we detect along some of the filamentary structures is SiO (see Fig. \ref{sio-others}). Figure \ref{sio-others} shows the detection of SiO along three filamentary structures: the eastern ridge, the filament located west of the protostar HOPS-64, and the central collimated filament. Its detection in the filamentary structures indicates that the gas is compressed and possibly shocked. The shocks could be induced by internal or external feedback processes. For example, the detection of narrow widespread SiO emission at parsec scale in high-mass star-forming filaments was thought to be caused by either gentle filament merging, that is, merging occurring at relatively low velocity \citep{jimenez-serra2010, Henshaw2013}, or by a decelerated shocked gas associated with large-scale outflows driven by the protostellar sources \citep{jimenez-serra2010, lopez-sepulcre2016}. Higher-resolution studies have added a new possibility, which is the collision between the star-forming cloud and the molecular gas pushed by H\,II regions or supernova remnants (e.g. \citealt{Cosentino2019,Cosentino2020}). Recently, both SiO and CH$_{3}$OH emission were observed at sub-parsec scales and $\sim$2$\arcsec$ angular resolution in the filamentary structures of IRAS\,4, a low-mass protostellar system located in the NGC 1333 star-forming region \citep{DeSimone2022}. The authors suggested that the emission arises from shock trains due to an expanding gas bubble, coming from behind NGC 1333 and clashing against the filament, where IRAS 4A lies. The narrow ($<$2 km\,s$^{-1}$) SiO HV imprint along the eastern ridge (see Appendix \ref{spectra}, Fig. \ref{sio-spec}) suggests that a similar scenario cannot be excluded in our case. Several molecular CO shells from B stars were observed in the surroundings of OMC-2\,FIR\,4 \citep{Feddersen2018}, in addition to an increase of CO emission at the eastern edge of the ISF \citep{Shimajiri2011}. Hence, the shocks in the filamentary structures of OMC-2\,FIR\,4 may be induced by the compression of the cloud due to some external mechanisms such as an expanding bubble that impact the protocluster or external winds from the OB stars located nearby. Alternatively, the SiO signature along the filamentary structures can be produced by the feedback of the protostars within the cluster, as discussed in the previous section. The coincidence of SiO with dust emission (see Appendix \ref{app}, Fig. \ref{sio-continuum}) could be indicative of dense or compressed gas that is not necessarily shocked. This is also consistent with the detection of the class I type CH$_{3}$OH masers along the eastern ridge. To comprehend which feedback process might be compressing the cluster and producing the SiO along the filamentary structures, we need to observe more molecular line transitions to derive the SiO abundance and the gas physical properties. An alternative interpretation of the SiO blueshifted emission along the eastern ridge is discussed in section \ref{outflow-hops-370}.

\subsection{SiO jet from VLA15}

\begin{figure}
    \centering
    \includegraphics[width=0.48\textwidth]{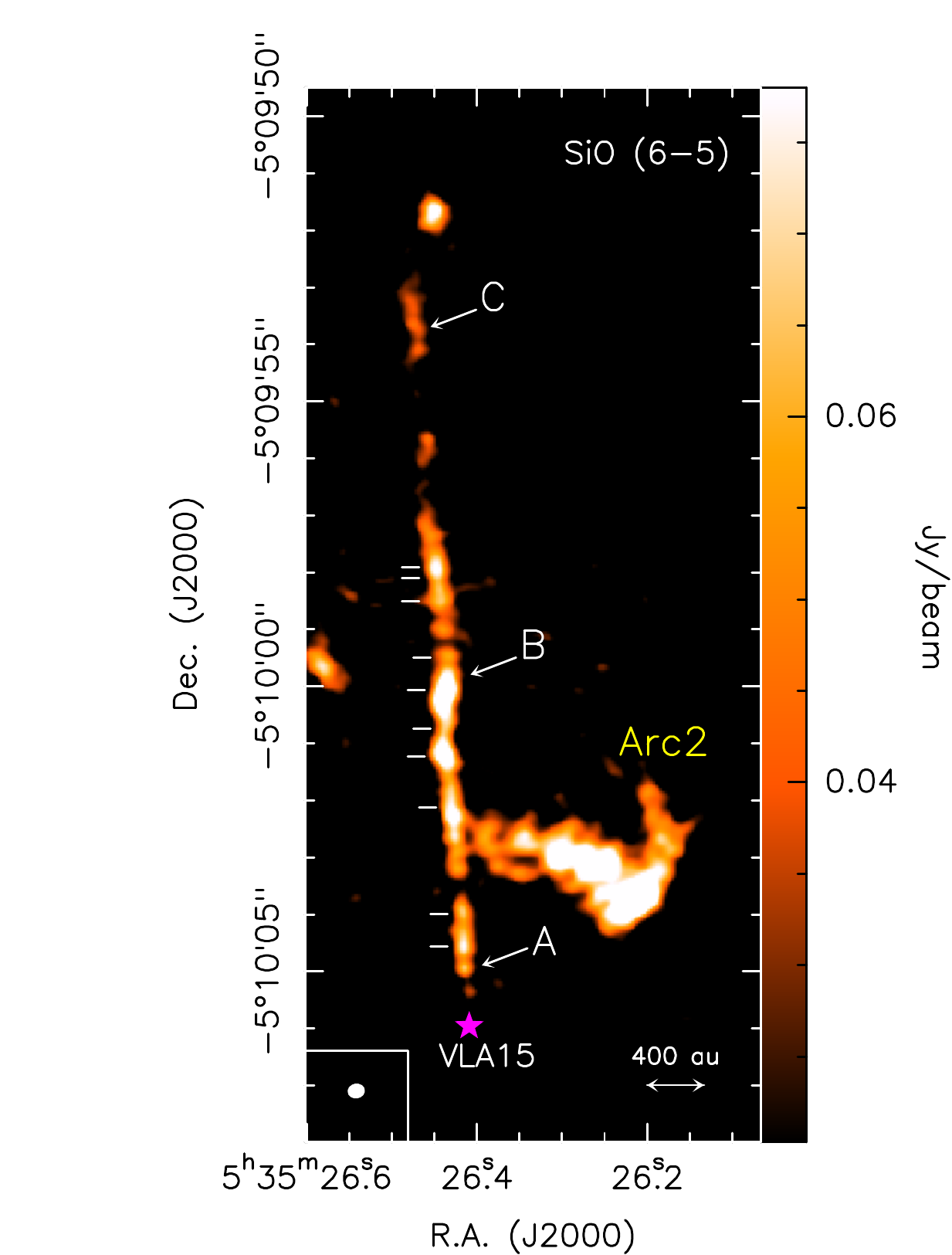}
    \caption{Velocity integrated emission of the SiO jet driven by VLA15. The colour images are for intensities higher than 3$\sigma$ with $\sigma$= 4.8 mJy beam$^{-1}$ km\,s$^{-1}$. The cut positions at which the widths were measured are shown with white horizontal lines. The knot positions at which the dynamical ages are estimated are shown with arrows and are labelled in white. VLA15 is depicted with a magenta star and labelled in white. The synthesised beam is depicted in the lower left corner.}
    \label{sio-jet}
\end{figure}

\begin{figure}
    \centering
    \includegraphics[width=0.45\textwidth]{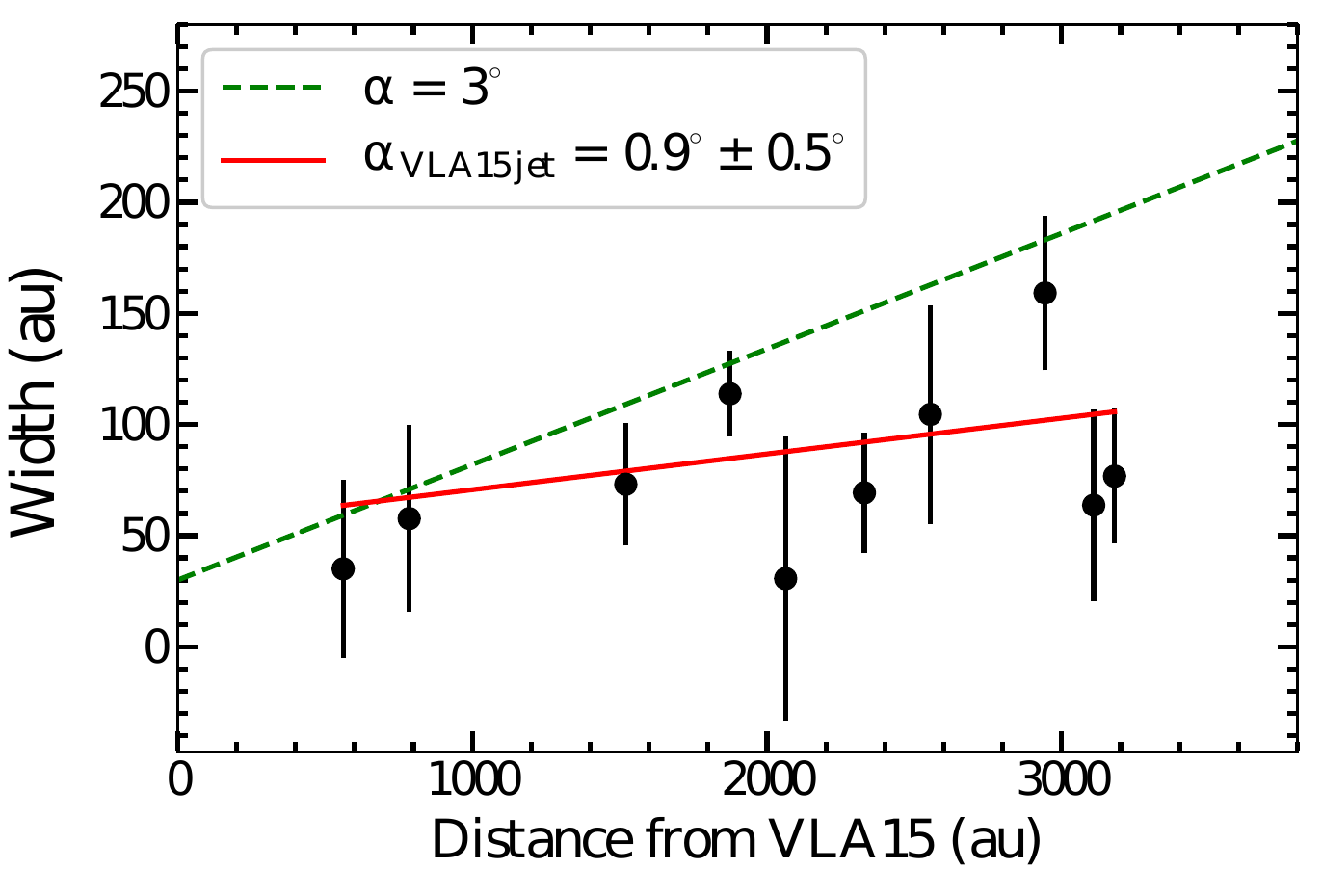}
    \caption{Deconvolved widths (2R$_{jet}$) of the SiO(6--5) emission along the jet axis as a function of the distance from VLA15. The solid red line corresponds to the best fit of the data points. The dashed green line corresponds to an opening angle of $\alpha$=3$^{\circ}$ and an initial width of 30 au.}
    \label{sio-cuts}
\end{figure}

Figure \ref{sio-jet} presents the SiO(6--5) clumpy elongated jet from VLA15. The jet is very narrow and shows a wiggle, similarly to what has been reported in HH212, another protostar located in Orion \citep{Lee2015,Lee2017}. To investigate the jet collimation, we measured the width along several cuts perpendicular to the jet axis where the SiO emission is $>$10$\sigma$. The widths were measured by fitting the spatial profile of each cut with a Gaussian profile. Then the deconvolved widths were derived by correcting the full width at half maximum (FWHM) of the observed emission for the size of the beam perpendicular to the jet axis employing the same equation as was used by \cite{Podio2021}. We plot the deconvolved sizes as a function of the distance from VLA15 in Fig. \ref{sio-cuts}. The width increases slightly with the distance. At 600-800 au from the source, the widths were estimated to be $\sim$35-60 au, slightly lower than the width of HH212 jet estimated at the same distances ($\sim$80 au; \citealt{Cabrit2007}). We derived the average collimation angle of the VLA15 jet by fitting the measured cut widths with a straight line, taking the measurement errors into account. We performed the fit using the Python function \textit{curve\_fit}, according to the following equation: 
\begin{equation}
    2R_{jet} = 2\,tan(\alpha/2)z +2R_0 ,
\end{equation}

\noindent where 2R$_{jet}$ is the deconvolved width, $z$ is the distance from the source VLA15, $\alpha$ is the apparent full opening angle of the flow\footnote{Due to projection effects, the true opening angle is smaller: tan$(\alpha_{\rm{true}})$= tan$(\alpha)\times$ sin$(i)$, where $i$ is the inclination with respect to the line of sight.} , and R$_{0}$ is a constant offset. For reference, we drew a dashed green line that represents a collimation with a full opening angle of $\alpha$=3$^{\circ}$ and initial width of 30 au, as was done by \cite{Podio2021}. Our estimated widths fall belowthis line. We derived an apparent full opening angle $\alpha$ = 0.9$^{\circ} \pm$0.5$^{\circ}$ from the fit, indicating that the jet of VLA15 is highly collimated ($<$2$^{\circ}$). The jet collimation agrees with the values derived for atomic jets from class II sources ($\sim$3$^{\circ}$), while it is smaller than those inferred for the class 0 SiO jets of the CALYPSO survey (4$^{\circ}$ - 12$^{\circ}$ at $>$300 au distances; \citealt{Podio2021}). Our value is also lower than the value inferred for HH212 IRAM-PdBI observations \citep{Codella2007, Cabrit2007}, possibly due to the lower spatial resolution ($\sim$150 au). The very small opening angle that we derive ($\sim$1$^{\circ}$) indicates that at the distance of the closest detected knot ($\sim$600 au) the jet is already collimated. Hence, our results suggest the VLA15 SiO jet is one of the most highly collimated jets observed so far. This is consistent with theoretical predictions that jets of young stellar objects are driven by magnetohydrodynamic (MHD) mechanisms \citep{Ray2021}. We verified that the narrow opening angle of the jet is not the result of interferometric filtering by modelling the interferometer response to simulated structures with widths ranging from 1/6 to 10 times the size of the synthesised beam. The widths of the structures were not narrowed, but rather conserved, indicating that they were not resolved out by the interferometer (see Appendix \ref{jet-sim}). Hence, the jet width we observe is real, as is the narrow opening angle we derived.

We estimated the dynamical age of three knots (A, B, and C; see Fig. \ref{sio-jet}) along the jet assuming again a jet velocity V$_{jet}$ = 100 km\,s$^{-1}$ \citep{Podio2021}. The disk associated with VLA15 appears to be nearly edge-on \citep{Osorio2017, Tobin2019}, hence its jet likely lies close to the plane of sky ($i$=0$^{\circ}$). In this case, the distances between VLA15 and knots A, B, and C would be $d_{(\rm A)} \sim$420 au, $d_{(\rm B)} \sim$2260 au, and $d_{(\rm C)} \sim$4795 au, respectively. Therefore, the dynamical ages would be $\tau_{(\rm A)} \sim$20 yr, $\tau_{(\rm B)} \sim$107 yr, and $\tau_{(\rm C)} \sim$227 yr. If the jet is inclined with respect to the plane of sky by 30$^{\circ}$ at most, the dynamical ages of the knots may be older than what was estimated assuming $i$=0$^{\circ}$ by 16\% at most.

We note that the radial velocity of the jet goes from V $\sim$ -6.2 km\,s$^{-1}$ to V$\sim$ +5.4 km\,s$^{-1}$ (see Fig. \ref{moms-1-all}). This might be due either to deceleration from an interaction with the material within the FIR\,4 cloud and/or to a change in jet direction with respect to the line of sight. The high-velocity dispersion, observed to the south of arc\,1 along the direction of the jet from VLA15, suggests that it could be interacting with this bow-shock.

\subsection{Possible role of the outflow from HOPS-370}
\label{outflow-hops-370}

To the north of the OMC-2\,FIR\,4 region lies the intermediate-mass protostar HOPS-370. This source has a rotating disk with an east-west velocity gradient. It also drives a high-velocity outflow going from -56 km\,s$^{-1}$ to 66 km\,s$^{-1}$ with respect to the HOPS-370 systemic velocity \citep{Tobin2019}. The outflow extends from the north-east (NE) to the south-west (SW) of HOPS-370, and it was observed down to $\sim$0.09 pc south of HOPS-370 \citep{Shimajiri2008}. It was identified through many outflow tracers such as CO \citep{Aso2000, Williams2003,Shimajiri2008,Shimajiri2015, Wu2005, Takahashi2008,Gonzalez-Garcia2016,Tobin2019}, HCO$^{+}$ \citep{Aso2000,Shimajiri2015}, CS, HCN \citep{Shimajiri2015}, and [O\,I] \citep{Gonzalez-Garcia2016}. It was also observed with VLA at 9 mm and 5 cm by \cite{Tobin2019} and \cite{Osorio2017}, respectively. The blue- and redshifted components of the outflow are observed in both directions \citep{Aso2000,Takahashi2008, Shimajiri2008, Tobin2019}. Some authors argued that the outflow directly interacts with clump FIR\,4 and triggers the star formation within it (e.g. \citealt{Shimajiri2008, Manoj2013, Shimajiri2015, Gonzalez-Garcia2016, Osorio2017}), while others excluded interaction and suggested that spatial coincidence between the outflow from HOPS-370 and clump FIR\,4 is a projection effect \citep{Favre2018}. 

The jet driven by HOPS-370 is likely precessing because the direction of the jet detected in [O\,I] by \cite{Gonzalez-Garcia2016} is not coincident with that of the 9 mm jet, nor with the axis of outflow cavities detected in CO, which are perpendicular to the disk of HOPS-370 \citep{Tobin2019}. This precession pattern is also supported in our work by the detection of two bright arcs seen in SiO emission, called arc\,1 and arc\,2. In Figs. \ref{moms-1-all}-\ref{moms-2-all} we draw arrows along the directions of the [O\,I] jet, the axis of the outflow cavities seen in CO, and the SiO peaks on the bow-shock features revealed by our observations.
The SiO moment 0 and 1 maps show multiple bow-shock features at different velocities, whose emission is not associated with that of the continuum, nor with that of the other tracers (see Figs. \ref{mom0-all} and \ref{sio-continuum}). The geometry of the bow shocks suggests that this is due to a jet activity from HOPS-370. More precisely, the bow shocks might be the signature of a precessing jet that extends from east to west. Hence, the western part of the protocluster is affected by the mass loss arriving from the north. Furthermore, SiO leaves a narrow-line imprint along the eastern ridge. This may be induced by one of the processes discussed in the previous section (i.e. gentle filament merging, winds from nearby H\,II regions, expanding shells, or stellar feedback). Alternatively, the jet from HOPS-370 may have also contributed to compress and shock the filamentary structures in the protocluster. Nonetheless, due to the limited number of molecular lines in our dataset, we were not able to prefer one of the two different scenarios.

In this paragraph, we estimate the dynamical ages of the arc\,1 and arc\,2 bow-shocks. The jet is thought to lie close to the plane of the sky \citep{Gonzalez-Garcia2016, Tobin2019}. When we consider the CO jet radial velocity ($\sim$50--60 km\,s$^{-1}$) derived by \cite{Tobin2019} and assume a typical jet velocity V$_{jet}$ = 100 km\,s$^{-1}$ \citep{Podio2021}, the inclination angle would be $i < 30^{\circ}$ with respect to the plane of the sky. If the jet lies on the plane of the sky ($i=0^{\circ}$), the distance between HOPS-370 and the SiO peak in arc\,1 and arc\,2 would be $d_{\rm (Arc\,1)} \sim$10330 au and $d_{\rm (Arc\,2)} \sim$14150 au. Hence, the dynamical ages would be $\tau_{\rm (Arc\,1)} = d_{\rm (Arc\,1)}$/ V$_{jet} \sim$490 yr and $\tau_{\rm (Arc\,2)} \sim$670 yr, respectively. The jet may be inclined with respect to the plane of the sky by $30^{\circ}$ at most, which implies that the shock dynamical ages may be older than what was estimated assuming $i$=0$^{\circ}$ by 15\% at most. The fact that arc\,2 is older than arc\,1 suggests that the jet of HOPS-370 is precessing from east to west and then proceeds away.

Interestingly, the outflow of HOPS-370 (producing arc\,2) does not appear to affect the highly collimated jet from VLA15. This indicates that the outflow and the jet are spatially separated, hence the proximity of arc\,2 to the VLA15 jet is merely a projection effect. This result, as well as the spatial dichotomy between the material traced in SiO and that traced by the other molecular tracers, confirms that the outflow from HOPS-370 has a tangential impact on OMC-2\,FIR\,4 rather than a direct one, as previously suggested by \cite{Tobin2019}.







\section{Conclusions}

\label{Conclusions}
We have investigated the internal structures of the protostellar cluster OMC-2\,FIR\,4 with ALMA observations at 0.04 pc scale and $\sim$100 au angular resolution. We mapped the emission of eight different molecular tracers: SiO, CH$_{3}$OH, C$^{18}$O, CS, CH$_{3}$CN, H$^{13}$CN, HC$_{3}$N, and CCH, probing high-density gas and shocks. The analysis was mainly focused on SiO and CH$_{3}$OH, and our main conclusions are summarised below.
\begin{enumerate}
    \item A net of intricate filamentary structures was revealed in all the tracers probing extended structures. One of the filamentary structures is aligned with one of the fibres observed in N$_{2}$H$^{+}$ by \cite{Hacar2018}, and the other short filamentary structures could be connected to the intra-hub fibres that are not resolved at parsec scales.
    
    \item From the analysis of CH$_{3}$OH kinematics, we showed that each filament moves at a different velocity. The filamentary structures show supersonic motions likely due to protostellar feedback.
    
    \item We observed SiO imprints on some filamentary structures suggesting that they trace compressed and/or shocked gas.
    
    \item SiO emission unveiled multiple bow-shock features with sizes between $\sim$500 and 2700 au. Their geometry and age suggest that they are likely caused by a precessing jet from HOPS-370 that extends from east to west. Their dynamical ages are $<$800 yr.
    
    \item We observed a spatial dichotomy between SiO and CH$_{3}$OH emission within the protocluster, where SiO dominates the western region and CH$_{3}$OH dominates the eastern region, suggesting that the jet from HOPS-370 does not directly impact OMC-2\,FIR\,4.
    
    \item We detected a highly collimated ($\sim$1$^{\circ}$) SiO jet extending along the S-N direction with a projected length of $\sim$5200 au from the embedded protostar VLA15.

\end{enumerate}
In a nutshell, the observations at high-angular resolutions and sub-parsec scales have allowed us to reveal the previously unresolved hub filamentary structures within OMC-2\,FIR\,4. Our study complements the studies performed at $\sim$1 pc scales using N$_{2}$H$^{+}$, and indicates that the formation of such complex structures could show how cluster-forming regions evolve, with the material being compressed within narrow filaments due to protostellar feedback. Imaging multiple molecular lines is hence important to characterise the gas properties and the processes at play within OMC-2\,FIR\,4 and other protoclusters.

\begin{acknowledgements}
This project has received funding from the European Union’s Horizon 2020 research and innovation program under the Marie Skłodowska-Curie grant agreement No 811312 for the Project "Astro-Chemical Origins” (ACO). This paper makes use of the following ALMA data: ADS/JAO.ALMA 2016.1.00376.S. ALMA is a partnership of ESO (representing its member states), NSF (USA) and NINS (Japan), together with NRC (Canada), MOST and ASIAA (Taiwan), and KASI (Republic of Korea), in cooperation with the Republic of Chile. The Joint ALMA Observatory is operated by ESO, AUI/NRAO and NAOJ. CCo and LP acknowledges the project PRIN-INAF 2016 The Cradle of Life - GENESIS-SKA (General Conditions in Early Planetary Systems for the rise of life with SKA), and the
PRIN-MUR 2020 MUR BEYOND-2p (Astrochemistry beyond the second period elements, Prot. 2020AFB3FX). ALS, CC, CCo, MDS and MB acknowledge the funding from the European Research Council (ERC) under the European Union’s Horizon 2020 research and innovation program, for the Project "The Dawn of Organic Chemistry" (DOC), grant agreement No 741002.  
\end{acknowledgements}

%
%

\bibliographystyle{aa} 
\bibliography{references}

\begin{thebibliography}{111}
\expandafter\ifx\csname natexlab\endcsname\relax\def\natexlab#1{#1}\fi

\bibitem[{{Adams}(2010)}]{Adams2010}
{Adams}, F.~C. 2010, \araa, 48, 47

\bibitem[{{Andr{\'e}} {et~al.}(2014){Andr{\'e}}, {Di Francesco},
  {Ward-Thompson}, {Inutsuka}, {Pudritz}, \& {Pineda}}]{Andre2014}
{Andr{\'e}}, P., {Di Francesco}, J., {Ward-Thompson}, D., {et~al.} 2014, in
  Protostars and Planets VI, ed. H.~{Beuther}, R.~S. {Klessen}, C.~P.
  {Dullemond}, \& T.~{Henning}, 27

\bibitem[{{Aso} {et~al.}(2000){Aso}, {Tatematsu}, {Sekimoto}, {Nakano},
  {Umemoto}, {Koyama}, \& {Yamamoto}}]{Aso2000}
{Aso}, Y., {Tatematsu}, K., {Sekimoto}, Y., {et~al.} 2000, \apjs, 131, 465

\bibitem[{{Bally} {et~al.}(1987){Bally}, {Langer}, {Stark}, \&
  {Wilson}}]{Bally1987}
{Bally}, J., {Langer}, W.~D., {Stark}, A.~A., \& {Wilson}, R.~W. 1987, \apjl,
  312, L45

\bibitem[{{Bouvier} {et~al.}(2020){Bouvier}, {L{\'o}pez-Sepulcre},
  {Ceccarelli}, {Kahane}, {Imai}, {Sakai}, {Yamamoto}, \&
  {Dagdigian}}]{Bouvier2020}
{Bouvier}, M., {L{\'o}pez-Sepulcre}, A., {Ceccarelli}, C., {et~al.} 2020, \aap,
  636, A19

\bibitem[{{Cabrit} {et~al.}(2007){Cabrit}, {Codella}, {Gueth}, {Nisini},
  {Gusdorf}, {Dougados}, \& {Bacciotti}}]{Cabrit2007}
{Cabrit}, S., {Codella}, C., {Gueth}, F., {et~al.} 2007, \aap, 468, L29

\bibitem[{{Caselli} {et~al.}(1997){Caselli}, {Hartquist}, \&
  {Havnes}}]{Caselli1997}
{Caselli}, P., {Hartquist}, T.~W., \& {Havnes}, O. 1997, \aap, 322, 296

\bibitem[{{Ceccarelli} {et~al.}(2014){Ceccarelli}, {Dominik},
  {L{\'o}pez-Sepulcre}, {Kama}, {Padovani}, {Caux}, \&
  {Caselli}}]{Ceccarelli2014}
{Ceccarelli}, C., {Dominik}, C., {L{\'o}pez-Sepulcre}, A., {et~al.} 2014,
  \apjl, 790, L1

\bibitem[{{Chahine} {et~al.}(2022){Chahine}, {L{\'o}pez-Sepulcre}, {Neri},
  {Ceccarelli}, {Mercimek}, {Codella}, {Bouvier}, {Bianchi}, {Favre}, {Podio},
  {Alves}, {Sakai}, \& {Yamamoto}}]{Chahine2022}
{Chahine}, L., {L{\'o}pez-Sepulcre}, A., {Neri}, R., {et~al.} 2022, \aap, 657,
  A78

\bibitem[{{Chen} {et~al.}(2019){Chen}, {Ellingsen}, {Ren}, {Sobolev},
  {Parfenov}, \& {Shen}}]{Chen2019}
{Chen}, X., {Ellingsen}, S.~P., {Ren}, Z.-Y., {et~al.} 2019, \apj, 877, 90

\bibitem[{{Chen} {et~al.}(2011){Chen}, {Ellingsen}, {Shen}, {Titmarsh}, \&
  {Gan}}]{Chen2011}
{Chen}, X., {Ellingsen}, S.~P., {Shen}, Z.-Q., {Titmarsh}, A., \& {Gan}, C.-G.
  2011, \apjs, 196, 9

\bibitem[{{Chini} {et~al.}(1997){Chini}, {Reipurth}, {Ward-Thompson}, {Bally},
  {Nyman}, {Sievers}, \& {Billawala}}]{Chini1997}
{Chini}, R., {Reipurth}, B., {Ward-Thompson}, D., {et~al.} 1997, \apjl, 474,
  L135

\bibitem[{{Chung} {et~al.}(2021){Chung}, {Lee}, {Kim}, {Gopinathan}, {Tafalla},
  {Caselli}, {Myers}, {Liu}, {Yoo}, {Kim}, {Kim}, {Soam}, {Cho}, {Kwon}, {Lee},
  \& {Kang}}]{Chung2021}
{Chung}, E.~J., {Lee}, C.~W., {Kim}, S., {et~al.} 2021, \apj, 919, 3

\bibitem[{{Clark}(1980)}]{Clark1980}
{Clark}, B.~G. 1980, \aap, 89, 377

\bibitem[{{Codella} {et~al.}(2007){Codella}, {Cabrit}, {Gueth}, {Cesaroni},
  {Bacciotti}, {Lefloch}, \& {McCaughrean}}]{Codella2007}
{Codella}, C., {Cabrit}, S., {Gueth}, F., {et~al.} 2007, \aap, 462, L53

\bibitem[{{Codella} {et~al.}(2014){Codella}, {Maury}, {Gueth}, {Maret},
  {Belloche}, {Cabrit}, \& {Andr{\'e}}}]{Codella2014}
{Codella}, C., {Maury}, A.~J., {Gueth}, F., {et~al.} 2014, \aap, 563, L3

\bibitem[{{Cosentino} {et~al.}(2019){Cosentino}, {Jim{\'e}nez-Serra},
  {Caselli}, {Henshaw}, {Barnes}, {Tan}, {Viti}, {Fontani}, \&
  {Wu}}]{Cosentino2019}
{Cosentino}, G., {Jim{\'e}nez-Serra}, I., {Caselli}, P., {et~al.} 2019, \apjl,
  881, L42

\bibitem[{{Cosentino} {et~al.}(2020){Cosentino}, {Jim{\'e}nez-Serra},
  {Henshaw}, {Caselli}, {Viti}, {Barnes}, {Tan}, {Fontani}, \&
  {Wu}}]{Cosentino2020}
{Cosentino}, G., {Jim{\'e}nez-Serra}, I., {Henshaw}, J.~D., {et~al.} 2020,
  \mnras, 499, 1666

\bibitem[{{Crimier} {et~al.}(2009){Crimier}, {Ceccarelli}, {Lefloch}, \&
  {Faure}}]{Crimier2009}
{Crimier}, N., {Ceccarelli}, C., {Lefloch}, B., \& {Faure}, A. 2009, \aap, 506,
  1229

\bibitem[{{Cuadrado} {et~al.}(2017){Cuadrado}, {Goicoechea}, {Cernicharo},
  {Fuente}, {Pety}, \& {Tercero}}]{Cuadrado2017}
{Cuadrado}, S., {Goicoechea}, J.~R., {Cernicharo}, J., {et~al.} 2017, \aap,
  603, A124

\bibitem[{{Dartois} {et~al.}(2019){Dartois}, {Chabot}, {Id Barkach}, {Rothard},
  {Aug{\'e}}, {Agnihotri}, {Domaracka}, \& {Boduch}}]{Dartois2019}
{Dartois}, E., {Chabot}, M., {Id Barkach}, T., {et~al.} 2019, \aap, 627, A55

\bibitem[{{De Simone} {et~al.}(2022){De Simone}, {Codella}, {Ceccarelli},
  {L{\'o}pez-Sepulcre}, {Neri}, {Rivera-Ortiz}, {Busquet}, {Caselli},
  {Bianchi}, {Fontani}, {Lefloch}, {Oya}, \& {Pineda}}]{DeSimone2022}
{De Simone}, M., {Codella}, C., {Ceccarelli}, C., {et~al.} 2022, \mnras
  [\eprint[arXiv]{2201.03434}]

\bibitem[{{Draine} {et~al.}(1983){Draine}, {Roberge}, \&
  {Dalgarno}}]{Draine1983}
{Draine}, B.~T., {Roberge}, W.~G., \& {Dalgarno}, A. 1983, \apj, 264, 485

\bibitem[{{Duley} \& {Williams}(1993)}]{Duley1993}
{Duley}, W.~W. \& {Williams}, D.~A. 1993, \mnras, 260, 37

\bibitem[{{Evans} {et~al.}(2022){Evans}, {Fontani}, {Vastel}, {Ceccarelli},
  {Caselli}, {L{\'o}pez-Sepulcre}, {Neri}, {Alves}, {Chahine}, {Favre}, \&
  {Lattanzi}}]{Evans2022}
{Evans}, L., {Fontani}, F., {Vastel}, C., {et~al.} 2022, \aap, 657, A136

\bibitem[{{Favre} {et~al.}(2018){Favre}, {Ceccarelli}, {L{\'o}pez-Sepulcre},
  {Fontani}, {Neri}, {Manigand}, {Kama}, {Caselli}, {Jaber Al-Edhari},
  {Kahane}, {Alves}, {Balucani}, {Bianchi}, {Caux}, {Codella}, {Dulieu},
  {Pineda}, {Sims}, \& {Theul{\'e}}}]{Favre2018}
{Favre}, C., {Ceccarelli}, C., {L{\'o}pez-Sepulcre}, A., {et~al.} 2018, \apj,
  859, 136

\bibitem[{{Feddersen} {et~al.}(2018){Feddersen}, {Arce}, {Kong}, {Shimajiri},
  {Nakamura}, {Hara}, {Ishii}, {Sasaki}, \& {Kawabe}}]{Feddersen2018}
{Feddersen}, J.~R., {Arce}, H.~G., {Kong}, S., {et~al.} 2018, \apj, 862, 121

\bibitem[{{Flower} \& {Pineau des Forets}(1994)}]{Flower1994}
{Flower}, D.~R. \& {Pineau des Forets}, G. 1994, \mnras, 268, 724

\bibitem[{{Fontani} {et~al.}(2017){Fontani}, {Ceccarelli}, {Favre}, {Caselli},
  {Neri}, {Sims}, {Kahane}, {Alves}, {Balucani}, {Bianchi}, {Caux}, {Jaber
  Al-Edhari}, {Lopez-Sepulcre}, {Pineda}, {Bachiller}, {Bizzocchi},
  {Bottinelli}, {Chacon-Tanarro}, {Choudhury}, {Codella}, {Coutens}, {Dulieu},
  {Feng}, {Rimola}, {Hily-Blant}, {Holdship}, {Jimenez-Serra}, {Laas},
  {Lefloch}, {Oya}, {Podio}, {Pon}, {Punanova}, {Quenard}, {Sakai}, {Spezzano},
  {Taquet}, {Testi}, {Theul{\'e}}, {Ugliengo}, {Vastel}, {Vasyunin}, {Viti},
  {Yamamoto}, \& {Wiesenfeld}}]{Fontani2017}
{Fontani}, F., {Ceccarelli}, C., {Favre}, C., {et~al.} 2017, \aap, 605, A57

\bibitem[{{Fontani} {et~al.}(2020){Fontani}, {Quaia}, {Ceccarelli}, {Colzi},
  {L{\'o}pez-Sepulcre}, {Favre}, {Kahane}, {Caselli}, {Codella}, {Podio}, \&
  {Viti}}]{Fontani2020}
{Fontani}, F., {Quaia}, G., {Ceccarelli}, C., {et~al.} 2020, \mnras, 493, 3412

\bibitem[{{Furlan} {et~al.}(2014){Furlan}, {Megeath}, {Osorio}, {Stutz},
  {Fischer}, {Ali}, {Stanke}, {Manoj}, {Adams}, \& {Tobin}}]{Furlan2014}
{Furlan}, E., {Megeath}, S.~T., {Osorio}, M., {et~al.} 2014, \apj, 786, 26

\bibitem[{{Goldsmith} {et~al.}(2008){Goldsmith}, {Heyer}, {Narayanan}, {Snell},
  {Li}, \& {Brunt}}]{Goldsmith2008}
{Goldsmith}, P.~F., {Heyer}, M., {Narayanan}, G., {et~al.} 2008, \apj, 680, 428

\bibitem[{{Gonz{\'a}lez-Garc{\'\i}a} {et~al.}(2016){Gonz{\'a}lez-Garc{\'\i}a},
  {Manoj}, {Watson}, {Vavrek}, {Megeath}, {Stutz}, {Osorio}, {Wyrowski},
  {Fischer}, {Tobin}, {S{\'a}nchez-Portal}, {Diaz Rodriguez}, \&
  {Wilson}}]{Gonzalez-Garcia2016}
{Gonz{\'a}lez-Garc{\'\i}a}, B., {Manoj}, P., {Watson}, D.~M., {et~al.} 2016,
  \aap, 596, A26

\bibitem[{{Gounelle} {et~al.}(2013){Gounelle}, {Chaussidon}, \&
  {Rollion-Bard}}]{Gounelle2013}
{Gounelle}, M., {Chaussidon}, M., \& {Rollion-Bard}, C. 2013, \apjl, 763, L33

\bibitem[{{Gratier} {et~al.}(2017){Gratier}, {Bron}, {Gerin}, {Pety}, {Guzman},
  {Orkisz}, {Bardeau}, {Goicoechea}, {Le Petit}, {Liszt}, {{\"O}berg},
  {Peretto}, {Roueff}, {Sievers}, \& {Tremblin}}]{Gratier2017}
{Gratier}, P., {Bron}, E., {Gerin}, M., {et~al.} 2017, \aap, 599, A100

\bibitem[{{Grossschedl} {et~al.}(2018){Grossschedl}, {Alves}, {Meingast},
  {Ackerl}, {Ascenso}, {Bouy}, {Burkert}, {Forbrich}, {Fuernkranz}, {Goodman},
  {Hacar}, {Herbst-Kiss}, {Lada}, {Larreina}, {Leschinski}, {Lombardi},
  {Moitinho}, {Mortimer}, \& {Zari}}]{Grossschedl2018}
{Grossschedl}, J.~E., {Alves}, J., {Meingast}, S., {et~al.} 2018, VizieR Online
  Data Catalog, J/A+A/619/A106

\bibitem[{{Guillet} {et~al.}(2011){Guillet}, {Pineau Des For{\^e}ts}, \&
  {Jones}}]{Guillet2011}
{Guillet}, V., {Pineau Des For{\^e}ts}, G., \& {Jones}, A.~P. 2011, \aap, 527,
  A123

\bibitem[{{Gusdorf} {et~al.}(2008{\natexlab{a}}){Gusdorf}, {Cabrit}, {Flower},
  \& {Pineau Des For{\^e}ts}}]{Gusdorf2008a}
{Gusdorf}, A., {Cabrit}, S., {Flower}, D.~R., \& {Pineau Des For{\^e}ts}, G.
  2008{\natexlab{a}}, \aap, 482, 809

\bibitem[{{Gusdorf} {et~al.}(2008{\natexlab{b}}){Gusdorf}, {Pineau Des
  For{\^e}ts}, {Cabrit}, \& {Flower}}]{Gusdorf2008b}
{Gusdorf}, A., {Pineau Des For{\^e}ts}, G., {Cabrit}, S., \& {Flower}, D.~R.
  2008{\natexlab{b}}, \aap, 490, 695

\bibitem[{{Hacar} {et~al.}(2017{\natexlab{a}}){Hacar}, {Alves}, {Tafalla}, \&
  {Goicoechea}}]{Hacar2017-ori}
{Hacar}, A., {Alves}, J., {Tafalla}, M., \& {Goicoechea}, J.~R.
  2017{\natexlab{a}}, \aap, 602, L2

\bibitem[{{Hacar} {et~al.}(2022){Hacar}, {Clark}, {Heitsch}, {Kainulainen},
  {Panopoulou}, {Seifried}, \& {Smith}}]{Hacar2022}
{Hacar}, A., {Clark}, S., {Heitsch}, F., {et~al.} 2022, arXiv e-prints,
  arXiv:2203.09562

\bibitem[{{Hacar} \& {Tafalla}(2011)}]{Hacar2011}
{Hacar}, A. \& {Tafalla}, M. 2011, \aap, 533, A34

\bibitem[{{Hacar} {et~al.}(2017{\natexlab{b}}){Hacar}, {Tafalla}, \&
  {Alves}}]{Hacar2017-ngc}
{Hacar}, A., {Tafalla}, M., \& {Alves}, J. 2017{\natexlab{b}}, \aap, 606, A123

\bibitem[{{Hacar} {et~al.}(2018){Hacar}, {Tafalla}, {Forbrich}, {Alves},
  {Meingast}, {Grossschedl}, \& {Teixeira}}]{Hacar2018}
{Hacar}, A., {Tafalla}, M., {Forbrich}, J., {et~al.} 2018, \aap, 610, A77

\bibitem[{{Hacar} {et~al.}(2013){Hacar}, {Tafalla}, {Kauffmann}, \&
  {Kov{\'a}cs}}]{Hacar2013}
{Hacar}, A., {Tafalla}, M., {Kauffmann}, J., \& {Kov{\'a}cs}, A. 2013, \aap,
  554, A55

\bibitem[{{Henshaw} {et~al.}(2014){Henshaw}, {Caselli}, {Fontani},
  {Jim{\'e}nez-Serra}, \& {Tan}}]{Henshaw2014}
{Henshaw}, J.~D., {Caselli}, P., {Fontani}, F., {Jim{\'e}nez-Serra}, I., \&
  {Tan}, J.~C. 2014, \mnras, 440, 2860

\bibitem[{{Henshaw} {et~al.}(2013){Henshaw}, {Caselli}, {Fontani},
  {Jim{\'e}nez-Serra}, {Tan}, \& {Hernandez}}]{Henshaw2013}
{Henshaw}, J.~D., {Caselli}, P., {Fontani}, F., {et~al.} 2013, \mnras, 428,
  3425

\bibitem[{{Henshaw} {et~al.}(2017){Henshaw}, {Jim{\'e}nez-Serra}, {Longmore},
  {Caselli}, {Pineda}, {Avison}, {Barnes}, {Tan}, \& {Fontani}}]{Henshaw2017}
{Henshaw}, J.~D., {Jim{\'e}nez-Serra}, I., {Longmore}, S.~N., {et~al.} 2017,
  \mnras, 464, L31

\bibitem[{{Hunter} {et~al.}(2014){Hunter}, {Brogan}, {Cyganowski}, \&
  {Young}}]{Hunter2014}
{Hunter}, T.~R., {Brogan}, C.~L., {Cyganowski}, C.~J., \& {Young}, K.~H. 2014,
  \apj, 788, 187

\bibitem[{{Jim{\'e}nez-Serra} {et~al.}(2010){Jim{\'e}nez-Serra}, {Caselli},
  {Tan}, {Hernandez}, {Fontani}, {Butler}, \& {van Loo}}]{jimenez-serra2010}
{Jim{\'e}nez-Serra}, I., {Caselli}, P., {Tan}, J.~C., {et~al.} 2010, \mnras,
  406, 187

\bibitem[{{Johnstone} \& {Bally}(1999)}]{Johnstone1999}
{Johnstone}, D. \& {Bally}, J. 1999, \apjl, 510, L49

\bibitem[{{Johnstone} {et~al.}(2003){Johnstone}, {Boonman}, \& {van
  Dishoeck}}]{Johnstone2003}
{Johnstone}, D., {Boonman}, A.~M.~S., \& {van Dishoeck}, E.~F. 2003, \aap, 412,
  157

\bibitem[{{Kainulainen} {et~al.}(2017){Kainulainen}, {Stutz}, {Stanke},
  {Abreu-Vicente}, {Beuther}, {Henning}, {Johnston}, \&
  {Megeath}}]{Kainulainen2017}
{Kainulainen}, J., {Stutz}, A.~M., {Stanke}, T., {et~al.} 2017, \aap, 600, A141

\bibitem[{{Kama} {et~al.}(2015){Kama}, {Caux}, {L{\'o}pez-Sepulcre}, {Wakelam},
  {Dominik}, {Ceccarelli}, {Lanza}, {Lique}, {Ochsendorf}, {Lis}, {Caballero},
  \& {Tielens}}]{Kama2015}
{Kama}, M., {Caux}, E., {L{\'o}pez-Sepulcre}, A., {et~al.} 2015, \aap, 574,
  A107

\bibitem[{{Kama} {et~al.}(2013){Kama}, {L{\'o}pez-Sepulcre}, {Dominik},
  {Ceccarelli}, {Fuente}, {Caux}, {Higgins}, {Tielens}, \&
  {Alonso-Albi}}]{Kama2013}
{Kama}, M., {L{\'o}pez-Sepulcre}, A., {Dominik}, C., {et~al.} 2013, \aap, 556,
  A57

\bibitem[{{Kang} {et~al.}(2013){Kang}, {Lee}, {Choi}, {Choi}, {Kim}, {Di
  Francesco}, \& {Park}}]{Kang2013}
{Kang}, M., {Lee}, J.-E., {Choi}, M., {et~al.} 2013, \apjs, 209, 25

\bibitem[{{Kogan} \& {Slysh}(1998)}]{Kogan1998}
{Kogan}, L. \& {Slysh}, V. 1998, \apj, 497, 800

\bibitem[{{Ladeyschikov} {et~al.}(2020){Ladeyschikov}, {Urquhart}, {Sobolev},
  {Breen}, \& {Bayandina}}]{Ladeyschikov2020}
{Ladeyschikov}, D.~A., {Urquhart}, J.~S., {Sobolev}, A.~M., {Breen}, S.~L., \&
  {Bayandina}, O.~S. 2020, \aj, 160, 213

\bibitem[{{Lattanzi} {et~al.}(2022){Lattanzi}, {Alves}, {Padovani}, \& {et
  al.}}]{Lattanzi2022}
{Lattanzi}, V., {Alves}, F., {Padovani}, M., \& {et al.} 2022, \aap, submitted

\bibitem[{{Lee} {et~al.}(2015){Lee}, {Hirano}, {Zhang}, {Shang}, {Ho}, \&
  {Mizuno}}]{Lee2015}
{Lee}, C.-F., {Hirano}, N., {Zhang}, Q., {et~al.} 2015, \apj, 805, 186

\bibitem[{{Lee} {et~al.}(2017){Lee}, {Ho}, {Li}, {Hirano}, {Zhang}, \&
  {Shang}}]{Lee2017}
{Lee}, C.-F., {Ho}, P. T.~P., {Li}, Z.-Y., {et~al.} 2017, Nature Astronomy, 1,
  0152

\bibitem[{{Leurini} {et~al.}(2010){Leurini}, {Parise}, {Schilke}, {Pety}, \&
  {Rolffs}}]{Leurini2010}
{Leurini}, S., {Parise}, B., {Schilke}, P., {Pety}, J., \& {Rolffs}, R. 2010,
  \aap, 511, A82

\bibitem[{{Leurini} {et~al.}(2004){Leurini}, {Schilke}, {Menten}, {Flower},
  {Pottage}, \& {Xu}}]{Leurini2004}
{Leurini}, S., {Schilke}, P., {Menten}, K.~M., {et~al.} 2004, \aap, 422, 573

\bibitem[{{Li} {et~al.}(2013){Li}, {Kauffmann}, {Zhang}, \& {Chen}}]{Li2013}
{Li}, D., {Kauffmann}, J., {Zhang}, Q., \& {Chen}, W. 2013, \apjl, 768, L5

\bibitem[{{L{\'o}pez-Sepulcre}
  {et~al.}(2013{\natexlab{a}}){L{\'o}pez-Sepulcre}, {Kama}, {Ceccarelli},
  {Dominik}, {Caux}, {Fuente}, \& {Alonso-Albi}}]{Loepz-Sepulcre-and-kama2013}
{L{\'o}pez-Sepulcre}, A., {Kama}, M., {Ceccarelli}, C., {et~al.}
  2013{\natexlab{a}}, \aap, 549, A114

\bibitem[{{L{\'o}pez-Sepulcre} {et~al.}(2017){L{\'o}pez-Sepulcre}, {Sakai},
  {Neri}, {Imai}, {Oya}, {Ceccarelli}, {Higuchi}, {Aikawa}, {Bottinelli},
  {Caux}, {Hirota}, {Kahane}, {Lefloch}, {Vastel}, {Watanabe}, \&
  {Yamamoto}}]{Lopez-Sepulcre2017}
{L{\'o}pez-Sepulcre}, A., {Sakai}, N., {Neri}, R., {et~al.} 2017, \aap, 606,
  A121

\bibitem[{{L{\'o}pez-Sepulcre}
  {et~al.}(2013{\natexlab{b}}){L{\'o}pez-Sepulcre}, {Taquet},
  {S{\'a}nchez-Monge}, {Ceccarelli}, {Dominik}, {Kama}, {Caux}, {Fontani},
  {Fuente}, {Ho}, {Neri}, \& {Shimajiri}}]{Lopez-Sepulcre-and-taquet2013}
{L{\'o}pez-Sepulcre}, A., {Taquet}, V., {S{\'a}nchez-Monge}, {\'A}., {et~al.}
  2013{\natexlab{b}}, \aap, 556, A62

\bibitem[{{L{\'o}pez-Sepulcre} {et~al.}(2016){L{\'o}pez-Sepulcre}, {Watanabe},
  {Sakai}, {Furuya}, {Saruwatari}, \& {Yamamoto}}]{lopez-sepulcre2016}
{L{\'o}pez-Sepulcre}, A., {Watanabe}, Y., {Sakai}, N., {et~al.} 2016, \apj,
  822, 85

\bibitem[{{Manoj} {et~al.}(2013){Manoj}, {Watson}, {Neufeld}, {Megeath},
  {Vavrek}, {Yu}, {Visser}, {Bergin}, {Fischer}, {Tobin}, {Stutz}, {Ali},
  {Wilson}, {Di Francesco}, {Osorio}, {Maret}, \& {Poteet}}]{Manoj2013}
{Manoj}, P., {Watson}, D.~M., {Neufeld}, D.~A., {et~al.} 2013, \apj, 763, 83

\bibitem[{{Maret} {et~al.}(2005){Maret}, {Ceccarelli}, {Tielens}, {Caux},
  {Lefloch}, {Faure}, {Castets}, \& {Flower}}]{Maret2005}
{Maret}, S., {Ceccarelli}, C., {Tielens}, A.~G.~G.~M., {et~al.} 2005, \aap,
  442, 527

\bibitem[{{Martin-Pintado} {et~al.}(1990){Martin-Pintado}, {Rodriguez-Franco},
  \& {Bachiller}}]{Martin-Pintado1990}
{Martin-Pintado}, J., {Rodriguez-Franco}, A., \& {Bachiller}, R. 1990, \apjl,
  357, L49

\bibitem[{{McMullin} {et~al.}(2007){McMullin}, {Waters}, {Schiebel}, {Young},
  \& {Golap}}]{McMullin2007}
{McMullin}, J.~P., {Waters}, B., {Schiebel}, D., {Young}, W., \& {Golap}, K.
  2007, in Astronomical Society of the Pacific Conference Series, Vol. 376,
  Astronomical Data Analysis Software and Systems XVI, ed. R.~A. {Shaw},
  F.~{Hill}, \& D.~J. {Bell}, 127

\bibitem[{{Mezger} {et~al.}(1990){Mezger}, {Wink}, \& {Zylka}}]{Mezger1990}
{Mezger}, P.~G., {Wink}, J.~E., \& {Zylka}, R. 1990, \aap, 228, 95

\bibitem[{{Minissale} {et~al.}(2016){Minissale}, {Dulieu}, {Cazaux}, \&
  {Hocuk}}]{Minissale2016}
{Minissale}, M., {Dulieu}, F., {Cazaux}, S., \& {Hocuk}, S. 2016, \aap, 585,
  A24

\bibitem[{{Molinari} {et~al.}(2014){Molinari}, {Bally}, {Glover}, {Moore},
  {Noriega-Crespo}, {Plume}, {Testi}, {V{\'a}zquez-Semadeni}, {Zavagno},
  {Bernard}, \& {Martin}}]{Molinari2014}
{Molinari}, S., {Bally}, J., {Glover}, S., {et~al.} 2014, in Protostars and
  Planets VI, ed. H.~{Beuther}, R.~S. {Klessen}, C.~P. {Dullemond}, \&
  T.~{Henning}, 125

\bibitem[{{M{\"u}ller} {et~al.}(2005){M{\"u}ller}, {Schl{\"o}der}, {Stutzki},
  \& {Winnewisser}}]{Muller2005}
{M{\"u}ller}, H. S.~P., {Schl{\"o}der}, F., {Stutzki}, J., \& {Winnewisser}, G.
  2005, Journal of Molecular Structure, 742, 215

\bibitem[{{M{\"u}ller} {et~al.}(2001){M{\"u}ller}, {Thorwirth}, {Roth}, \&
  {Winnewisser}}]{Muller2001}
{M{\"u}ller}, H.~S.~P., {Thorwirth}, S., {Roth}, D.~A., \& {Winnewisser}, G.
  2001, \aap, 370, L49

\bibitem[{{Myers}(1983)}]{Myers1983}
{Myers}, P.~C. 1983, \apj, 270, 105

\bibitem[{{Myers}(2009)}]{Myers2009}
{Myers}, P.~C. 2009, \apj, 700, 1609

\bibitem[{{Nesterenok}(2022)}]{Nesterenok2022}
{Nesterenok}, A.~V. 2022, \mnras, 509, 4555

\bibitem[{{Orkisz} {et~al.}(2019){Orkisz}, {Peretto}, {Pety}, {Gerin},
  {Levrier}, {Bron}, {Bardeau}, {Goicoechea}, {Gratier}, {Guzm{\'a}n},
  {Hughes}, {Languignon}, {Le Petit}, {Liszt}, {{\"O}berg}, {Roueff},
  {Sievers}, \& {Tremblin}}]{Orkisz2019}
{Orkisz}, J.~H., {Peretto}, N., {Pety}, J., {et~al.} 2019, \aap, 624, A113

\bibitem[{{Osorio} {et~al.}(2017){Osorio}, {D{\'\i}az-Rodr{\'\i}guez},
  {Anglada}, {Megeath}, {Rodr{\'\i}guez}, {Tobin}, {Stutz}, {Furlan},
  {Fischer}, {Manoj}, {G{\'o}mez}, {Gonz{\'a}lez-Garc{\'\i}a}, {Stanke},
  {Watson}, {Loinard}, {Vavrek}, \& {Carrasco-Gonz{\'a}lez}}]{Osorio2017}
{Osorio}, M., {D{\'\i}az-Rodr{\'\i}guez}, A.~K., {Anglada}, G., {et~al.} 2017,
  \apj, 840, 36

\bibitem[{{Pfalzner} {et~al.}(2015){Pfalzner}, {Davies}, {Gounelle},
  {Johansen}, {M{\"u}nker}, {Lacerda}, {Portegies Zwart}, {Testi}, {Trieloff},
  \& {Veras}}]{Pfalzner2015}
{Pfalzner}, S., {Davies}, M.~B., {Gounelle}, M., {et~al.} 2015, \physscr, 90,
  068001

\bibitem[{{Pihlstr{\"o}m} {et~al.}(2014){Pihlstr{\"o}m}, {Sjouwerman}, {Frail},
  {Claussen}, {Mesler}, \& {McEwen}}]{Pihlstrom2014}
{Pihlstr{\"o}m}, Y.~M., {Sjouwerman}, L.~O., {Frail}, D.~A., {et~al.} 2014,
  \aj, 147, 73

\bibitem[{{Podio} {et~al.}(2021){Podio}, {Tabone}, {Codella}, {Gueth}, {Maury},
  {Cabrit}, {Lefloch}, {Maret}, {Belloche}, {Andr{\'e}}, {Anderl}, {Gaudel}, \&
  {Testi}}]{Podio2021}
{Podio}, L., {Tabone}, B., {Codella}, C., {et~al.} 2021, \aap, 648, A45

\bibitem[{{Polychroni} {et~al.}(2013){Polychroni}, {Schisano}, {Elia}, {Roy},
  {Molinari}, {Martin}, {Andr{\'e}}, {Turrini}, {Rygl}, {Di Francesco},
  {Benedettini}, {Busquet}, {di Giorgio}, {Pestalozzi}, {Pezzuto},
  {Arzoumanian}, {Bontemps}, {Hennemann}, {Hill}, {K{\"o}nyves},
  {Men'shchikov}, {Motte}, {Nguyen-Luong}, {Peretto}, {Schneider}, \&
  {White}}]{Polychroni2013}
{Polychroni}, D., {Schisano}, E., {Elia}, D., {et~al.} 2013, \apjl, 777, L33

\bibitem[{{Ray} \& {Ferreira}(2021)}]{Ray2021}
{Ray}, T.~P. \& {Ferreira}, J. 2021, \nar, 93, 101615

\bibitem[{{Rimola} {et~al.}(2014){Rimola}, {Taquet}, {Ugliengo}, {Balucani}, \&
  {Ceccarelli}}]{Rimola2014}
{Rimola}, A., {Taquet}, V., {Ugliengo}, P., {Balucani}, N., \& {Ceccarelli}, C.
  2014, \aap, 572, A70

\bibitem[{{Rodriguez-Franco} {et~al.}(1992){Rodriguez-Franco},
  {Martin-Pintado}, {Gomez-Gonzalez}, \& {Planesas}}]{Rodriguez-Franco1992}
{Rodriguez-Franco}, A., {Martin-Pintado}, J., {Gomez-Gonzalez}, J., \&
  {Planesas}, P. 1992, \aap, 264, 592

\bibitem[{{Schilke} {et~al.}(1997){Schilke}, {Walmsley}, {Pineau des Forets},
  \& {Flower}}]{Schilke1997}
{Schilke}, P., {Walmsley}, C.~M., {Pineau des Forets}, G., \& {Flower}, D.~R.
  1997, \aap, 321, 293

\bibitem[{{Sch{\"o}ier} {et~al.}(2005){Sch{\"o}ier}, {van der Tak}, {van
  Dishoeck}, \& {Black}}]{Schoier2005}
{Sch{\"o}ier}, F.~L., {van der Tak}, F.~F.~S., {van Dishoeck}, E.~F., \&
  {Black}, J.~H. 2005, \aap, 432, 369

\bibitem[{{Shimajiri} {et~al.}(2011){Shimajiri}, {Kawabe}, {Takakuwa}, {Saito},
  {Tsukagoshi}, {Momose}, {Ikeda}, {Akiyama}, {Austermann}, {Ezawa}, {Fukue},
  {Hiramatsu}, {Hughes}, {Kitamura}, {Kohno}, {Kurono}, {Scott}, {Wilson},
  {Yoshida}, \& {Yun}}]{Shimajiri2011}
{Shimajiri}, Y., {Kawabe}, R., {Takakuwa}, S., {et~al.} 2011, \pasj, 63, 105

\bibitem[{{Shimajiri} {et~al.}(2015){Shimajiri}, {Sakai}, {Kitamura},
  {Tsukagoshi}, {Saito}, {Nakamura}, {Momose}, {Takakuwa}, {Yamaguchi},
  {Sakai}, {Yamamoto}, \& {Kawabe}}]{Shimajiri2015}
{Shimajiri}, Y., {Sakai}, T., {Kitamura}, Y., {et~al.} 2015, \apjs, 221, 31

\bibitem[{{Shimajiri} {et~al.}(2008){Shimajiri}, {Takahashi}, {Takakuwa},
  {Saito}, \& {Kawabe}}]{Shimajiri2008}
{Shimajiri}, Y., {Takahashi}, S., {Takakuwa}, S., {Saito}, M., \& {Kawabe}, R.
  2008, \apj, 683, 255

\bibitem[{{Shirley}(2015)}]{Shirley2015}
{Shirley}, Y.~L. 2015, \pasp, 127, 299

\bibitem[{{Sobolev} {et~al.}(2007){Sobolev}, {Cragg}, {Ellingsen}, {Gaylard},
  {Goedhart}, {Henkel}, {Kirsanova}, {Ostrovskii}, {Pankratova}, {Shelemei},
  {van der Walt}, {Vasyunina}, \& {Voronkov}}]{Sobolev2007}
{Sobolev}, A.~M., {Cragg}, D.~M., {Ellingsen}, S.~P., {et~al.} 2007, in
  Astrophysical Masers and their Environments, ed. J.~M. {Chapman} \& W.~A.
  {Baan}, Vol. 242, 81--88

\bibitem[{{Sokolov} {et~al.}(2019){Sokolov}, {Wang}, {Pineda}, {Caselli},
  {Henshaw}, {Barnes}, {Tan}, {Fontani}, \& {Jim{\'e}nez-Serra}}]{Sokolov2019}
{Sokolov}, V., {Wang}, K., {Pineda}, J.~E., {et~al.} 2019, \apj, 872, 30

\bibitem[{{Suri} {et~al.}(2019){Suri}, {S{\'a}nchez-Monge}, {Schilke},
  {Clarke}, {Smith}, {Ossenkopf-Okada}, {Klessen}, {Padoan}, {Goldsmith},
  {Arce}, {Bally}, {Carpenter}, {Ginsburg}, {Johnstone}, {Kauffmann}, {Kong},
  {Lis}, {Mairs}, {Pillai}, {Pineda}, \& {Duarte-Cabral}}]{Suri2019}
{Suri}, S., {S{\'a}nchez-Monge}, {\'A}., {Schilke}, P., {et~al.} 2019, \aap,
  623, A142

\bibitem[{{Tafalla} \& {Hacar}(2015)}]{Tafalla2015}
{Tafalla}, M. \& {Hacar}, A. 2015, \aap, 574, A104

\bibitem[{{Takahashi} {et~al.}(2008){Takahashi}, {Saito}, {Ohashi}, {Kusakabe},
  {Takakuwa}, {Shimajiri}, {Tamura}, \& {Kawabe}}]{Takahashi2008}
{Takahashi}, S., {Saito}, M., {Ohashi}, N., {et~al.} 2008, \apj, 688, 344

\bibitem[{{Tatematsu} {et~al.}(2008){Tatematsu}, {Kandori}, {Umemoto}, \&
  {Sekimoto}}]{Tatematsu2008}
{Tatematsu}, K., {Kandori}, R., {Umemoto}, T., \& {Sekimoto}, Y. 2008, \pasj,
  60, 407

\bibitem[{{Tielens} \& {Hagen}(1982)}]{Tielens1982}
{Tielens}, A.~G.~G.~M. \& {Hagen}, W. 1982, \aap, 114, 245

\bibitem[{{Tobin} {et~al.}(2019){Tobin}, {Megeath}, {van't Hoff},
  {D{\'\i}az-Rodr{\'\i}guez}, {Reynolds}, {Osorio}, {Anglada}, {Furlan},
  {Karnath}, {Offner}, {Sheehan}, {Sadavoy}, {Stutz}, {Fischer}, {Kama},
  {Persson}, {Di Francesco}, {Looney}, {Watson}, {Li}, {Stephens}, {Chandler},
  {Cox}, {Dunham}, {Kratter}, {Kounkel}, {Mazur}, {Murillo}, {Patel}, {Perez},
  {Segura-Cox}, {Sharma}, {Tychoniec}, \& {Wyrowski}}]{Tobin2019}
{Tobin}, J.~J., {Megeath}, S.~T., {van't Hoff}, M., {et~al.} 2019, \apj, 886, 6

\bibitem[{{Voronkov} {et~al.}(2006){Voronkov}, {Brooks}, {Sobolev},
  {Ellingsen}, {Ostrovskii}, \& {Caswell}}]{Voronkov2006}
{Voronkov}, M.~A., {Brooks}, K.~J., {Sobolev}, A.~M., {et~al.} 2006, \mnras,
  373, 411

\bibitem[{{Voronkov} {et~al.}(2014){Voronkov}, {Caswell}, {Ellingsen}, {Green},
  \& {Breen}}]{Voronkov2014}
{Voronkov}, M.~A., {Caswell}, J.~L., {Ellingsen}, S.~P., {Green}, J.~A., \&
  {Breen}, S.~L. 2014, \mnras, 439, 2584

\bibitem[{{Voronkov} {et~al.}(2010){Voronkov}, {Caswell}, {Ellingsen}, \&
  {Sobolev}}]{Voronkov2010}
{Voronkov}, M.~A., {Caswell}, J.~L., {Ellingsen}, S.~P., \& {Sobolev}, A.~M.
  2010, \mnras, 405, 2471

\bibitem[{{Watanabe} \& {Kouchi}(2002)}]{Watanabe2002}
{Watanabe}, N. \& {Kouchi}, A. 2002, \apjl, 571, L173

\bibitem[{{Williams} {et~al.}(2003){Williams}, {Plambeck}, \&
  {Heyer}}]{Williams2003}
{Williams}, J.~P., {Plambeck}, R.~L., \& {Heyer}, M.~H. 2003, \apj, 591, 1025

\bibitem[{{Wu} {et~al.}(2018){Wu}, {Qiu}, {Esimbek}, {Zheng}, {Henkel}, {Li},
  \& {Han}}]{Wu2018}
{Wu}, G., {Qiu}, K., {Esimbek}, J., {et~al.} 2018, \aap, 616, A111

\bibitem[{{Wu} {et~al.}(2005){Wu}, {Zhang}, {Chen}, {Yang}, {Wei}, \&
  {Ho}}]{Wu2005}
{Wu}, Y., {Zhang}, Q., {Chen}, H., {et~al.} 2005, \aj, 129, 330

\bibitem[{{Zhang} {et~al.}(2020){Zhang}, {Ren}, {Wu}, {Li}, {Zhu}, {Zhang},
  {Mardones}, {Wang}, {Shi}, {Yue}, {Luo}, {Xie}, {Jiao}, {Liu}, {Xu}, \&
  {Wang}}]{Zhang2020}
{Zhang}, C., {Ren}, Z., {Wu}, J., {et~al.} 2020, \mnras, 497, 793

\end{thebibliography}

\begin{appendix}
\onecolumn

\section{Spectra}

\label{spectra}

\begin{figure*}[hbt]
    \centering
    \includegraphics[width=\textwidth]{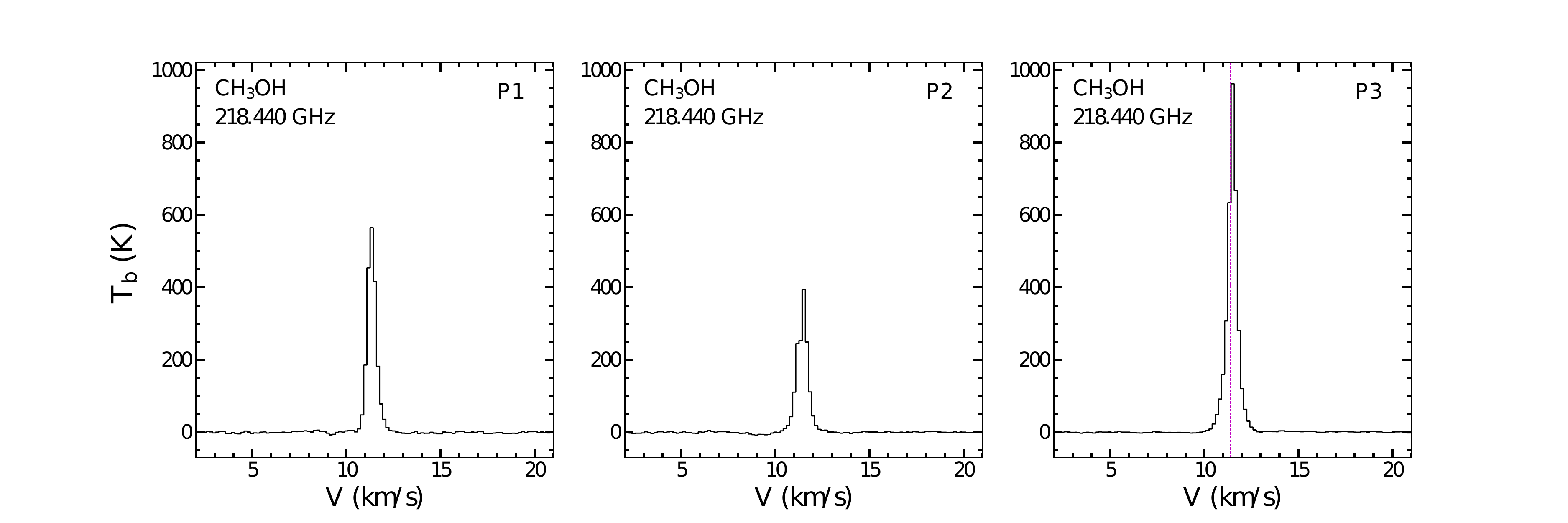}
    \caption{Spectra of the masers detected in CH$_{3}$OH(4$_{-2,3}-$3$_{-1,2}$\,E) at the P1-P3 positions. The vertical dashed line stands for the systemic velocity of OMC-2\,FIR\,4}
    \label{masers-spec}
\end{figure*}

\begin{figure}[hbt]
    \centering
    \includegraphics[width=\textwidth]{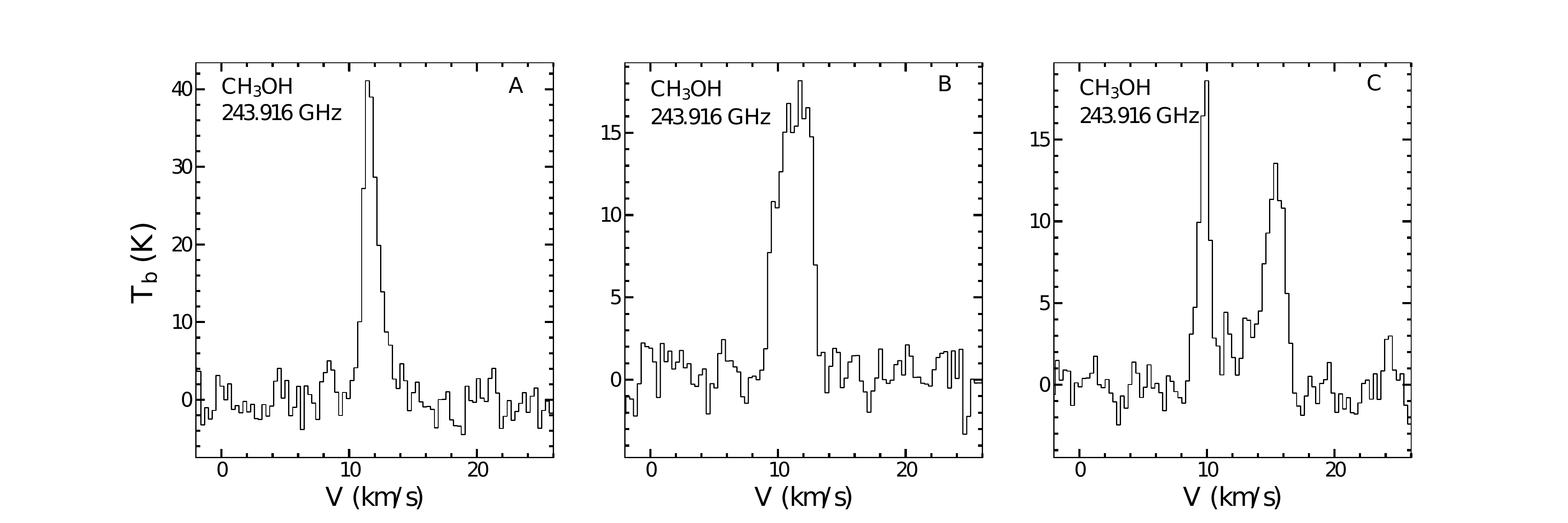}
    \caption{  CH$_{3}$OH (5$_{1,4}-$4$_{1,3}$\,A) spectra extracted from three different positions (A, B, and C) shown in the moment-2 map (See Fig. \ref{moms-2-all}). The first shows a region with low-velocity dispersion, the second shows a region with intermediate-velocity dispersion, and the third shows the central region with several components along the line of sight.}
    \label{meth-spec}
\end{figure}

\begin{figure}[hbt]

   \begin{minipage}{0.48\textwidth}
       
    \includegraphics[width=0.48\textwidth]{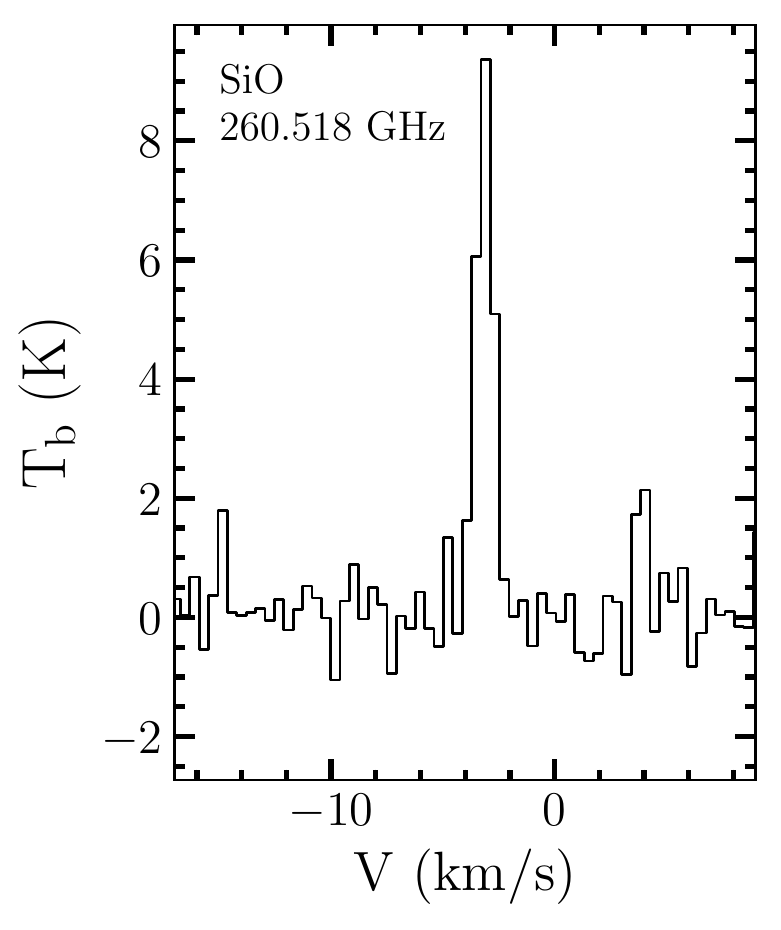}
    \caption{SiO emission representative of the eastern ridge extracted at the position (6.5$\arcsec$,2.1$\arcsec$)}
    \label{sio-spec}
    \end{minipage}\hfill
\end{figure}

We report in Fig. \ref{masers-spec} the spectra of CH$_{3}$OH (4$_{-2,3}-$3$_{-1,2}$\,E) masers, observed at three spots along the eastern ridge of OMC-2\,FIR\,4. In Fig. \ref{meth-spec} we show the CH$_{3}$OH (5$_{1,4}-$4$_{1,3}$\,A) spectra extracted from the different positions shown Fig. \ref{moms-2-all}. In Fig. \ref{sio-spec} we present an SiO spectrum extracted at a position along the eastern ridge.

\newpage
\section{Jet simulation}

\label{jet-sim}

\begin{figure}[h!]
    \centering
    \includegraphics[width=0.9\textwidth]{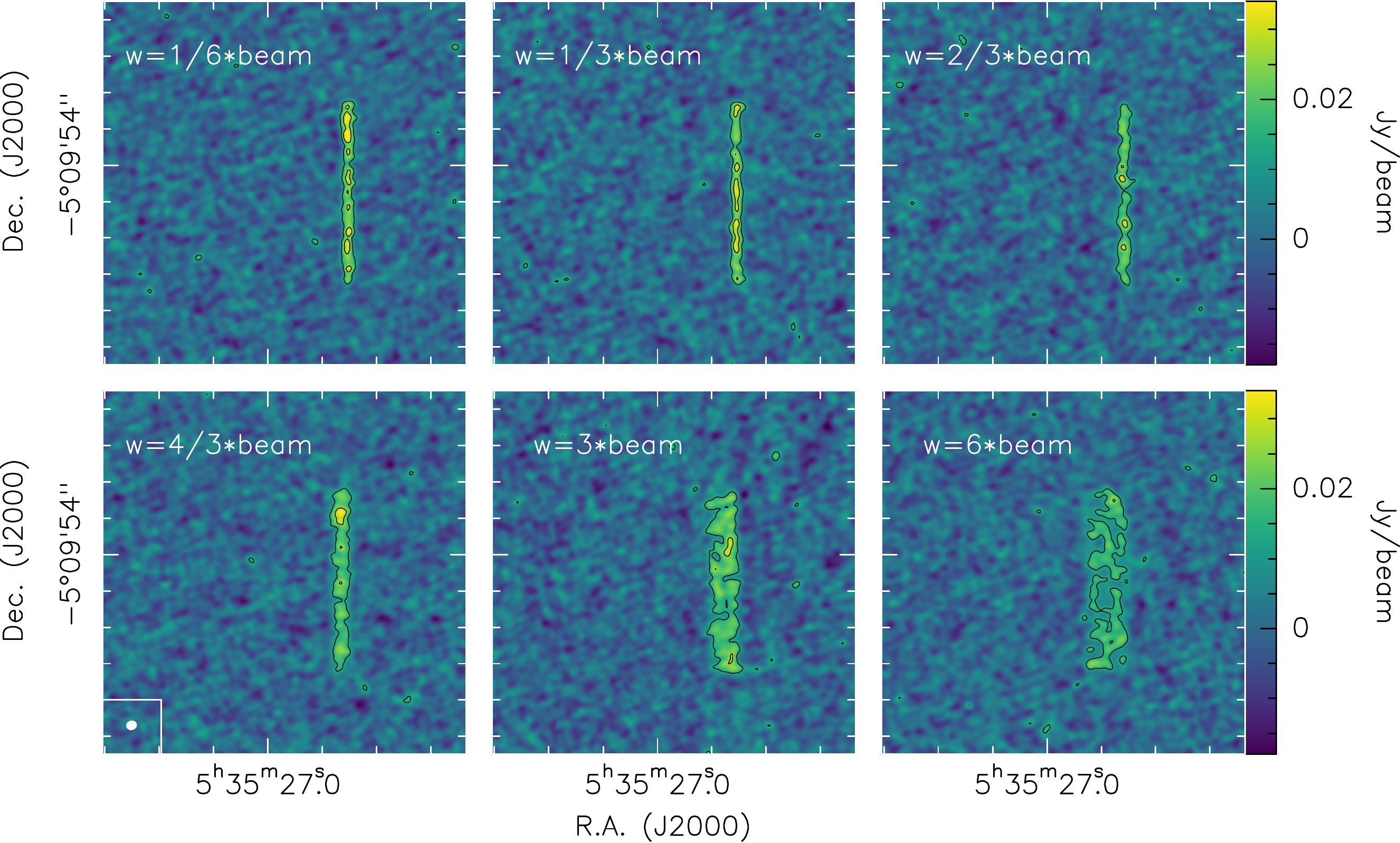}
    \caption{ Interferometric response to each structure after running the model. The width w of the rectangular (jet-like) structures given as input in the model is depicted in white in the upper left corner. The synthesised beam is depicted in white in the lower left corner. The colour scale is the same for all the panels. The box size is 5$" \times$5$"$.}
    \label{jet-sim-fig}
\end{figure}

In this section, we describe the simulation that we have performed to verify that the narrow opening angle of the jet is not the result of interferometric filtering. We first created rectangular structures with a fixed length and different widths between 1/6 and 10 times the size of the synthesised beam ($\sim$0$\farcs$05--3$\farcs$2). The interferometric response to each structure was determined by entering the visibilities corresponding to each structure into the SiO jet uv-table. The visibility weights were kept unchanged. To investigate the influence of thermal noise on the image reconstruction, Gaussian noise was added to the visibilities according to their weights. Finally, the uv-table of each structure was Fourier-transformed to produce a dirty map, and was then cleaned. The results are shown for a possible thermal noise model in Fig. \ref{jet-sim-fig}.



\newpage
\section{Maps}

\label{app}

\begin{figure*}[h!]
    \centering
\centering
   \includegraphics[width=0.48\textwidth]{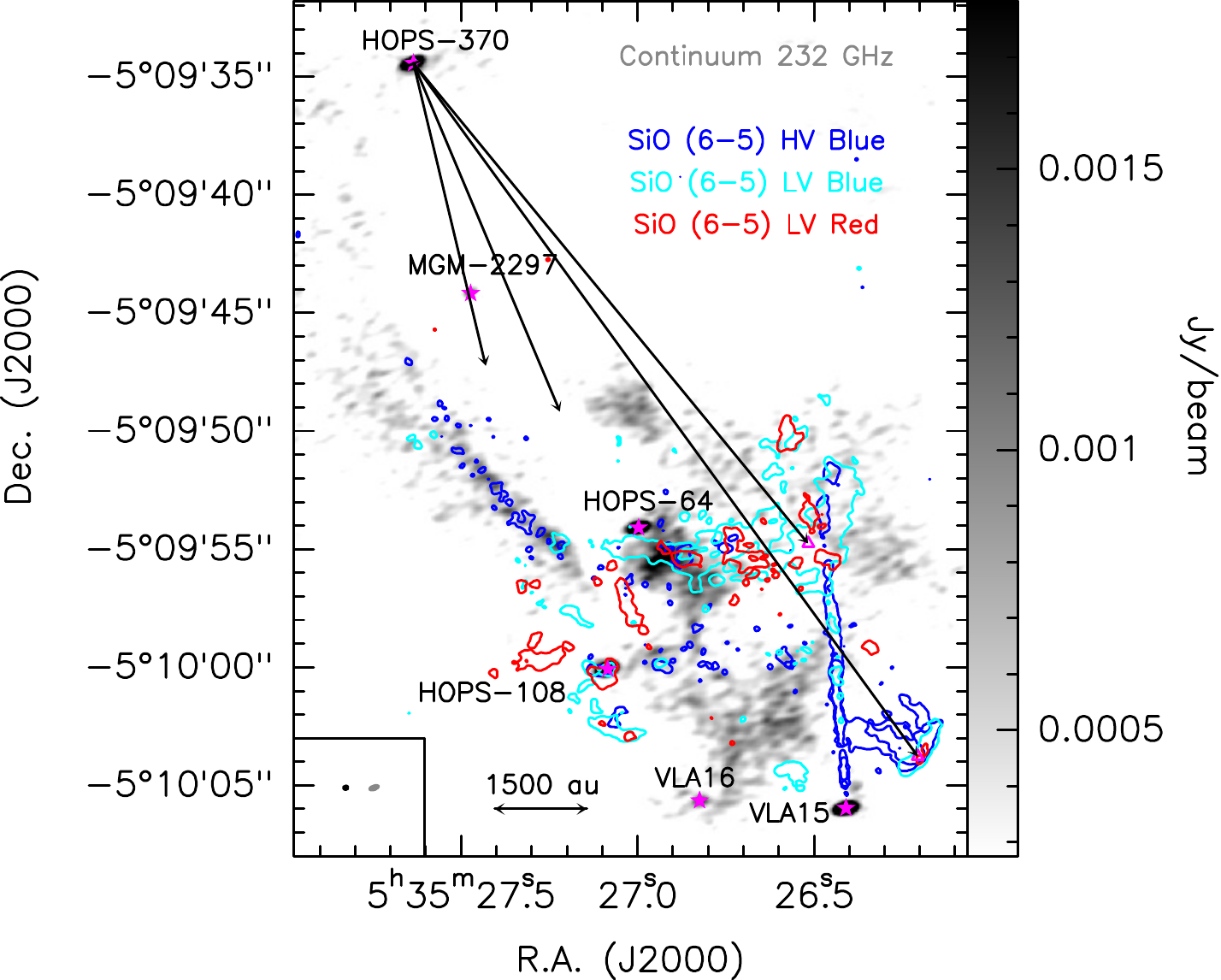}
    \caption{Continuum emission at 232 GHz in colours, with superimposed contours of the velocity-integrated emission of SiO at different velocity regimes. The colour images are for intensities higher than 3$\sigma$ with $\sigma$ = 91.7 $\mu$Jy beam$^{-1}$. The SiO contours are at 5$\sigma$ with $\sigma$= 5.8 mJy\,beam$^{-1}$\,km\,s$^{-1}$. They correspond to the blueshifted high-velocity regime (blue), the blueshifted low-velocity regime (cyan), and the redshifted low-velocity regime (red) (See Sect. \ref{kinematics}). The different cores are depicted with magenta stars and are labelled in black. The positions of the SiO peaks from this study are depicted with open magenta triangles. We draw four arrows from HOPS-370, one perpendicular to the disk, two towards the SiO peaks, and one similar to the direction of the jet in [O\,I] by \cite{Gonzalez-Garcia2016}. The synthesised beam of SiO and continuum are depicted in the lower left corner in blue and grey, respectively. }
   \label{sio-continuum}
\end{figure*}

\begin{figure*}[!]
    \centering
    \includegraphics[width=\textwidth]{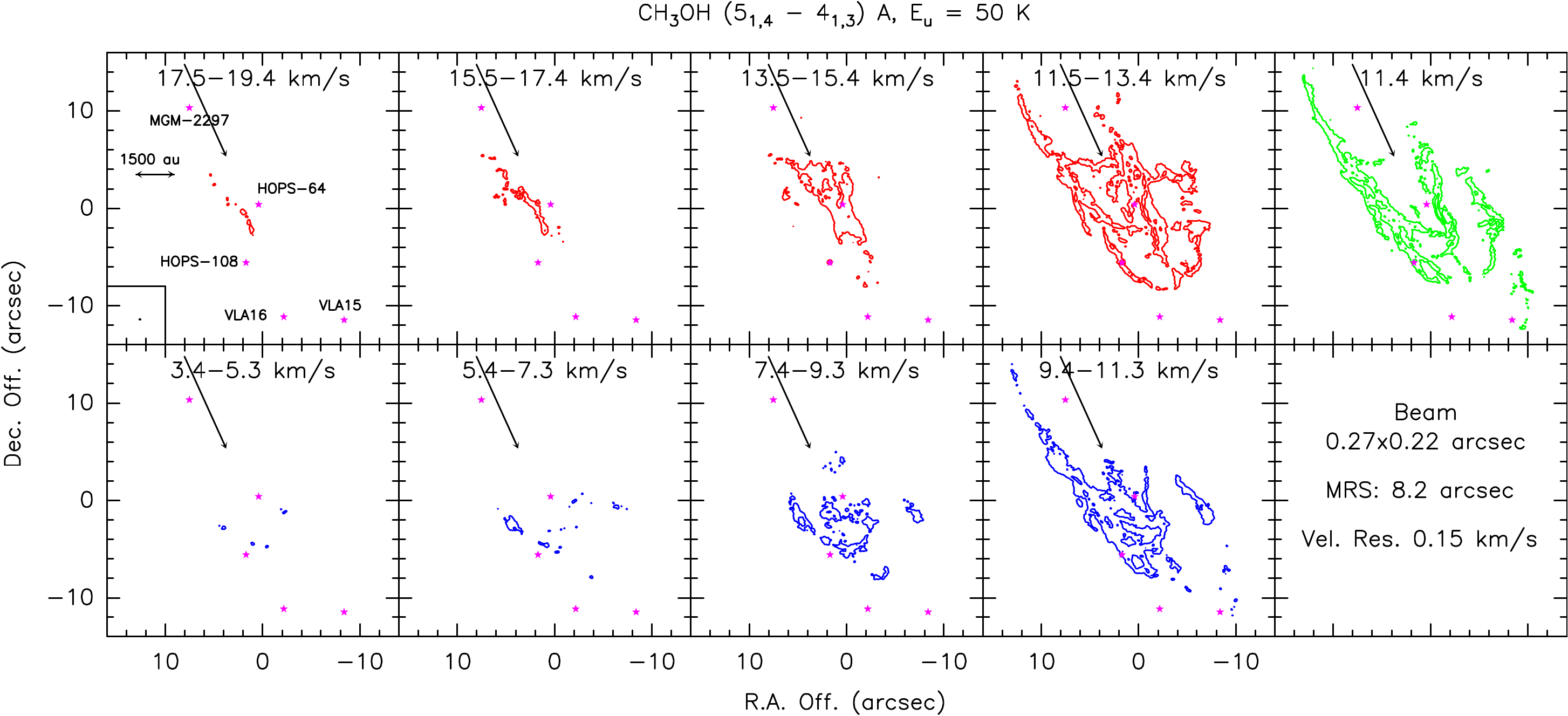}
    \includegraphics[width=\textwidth]{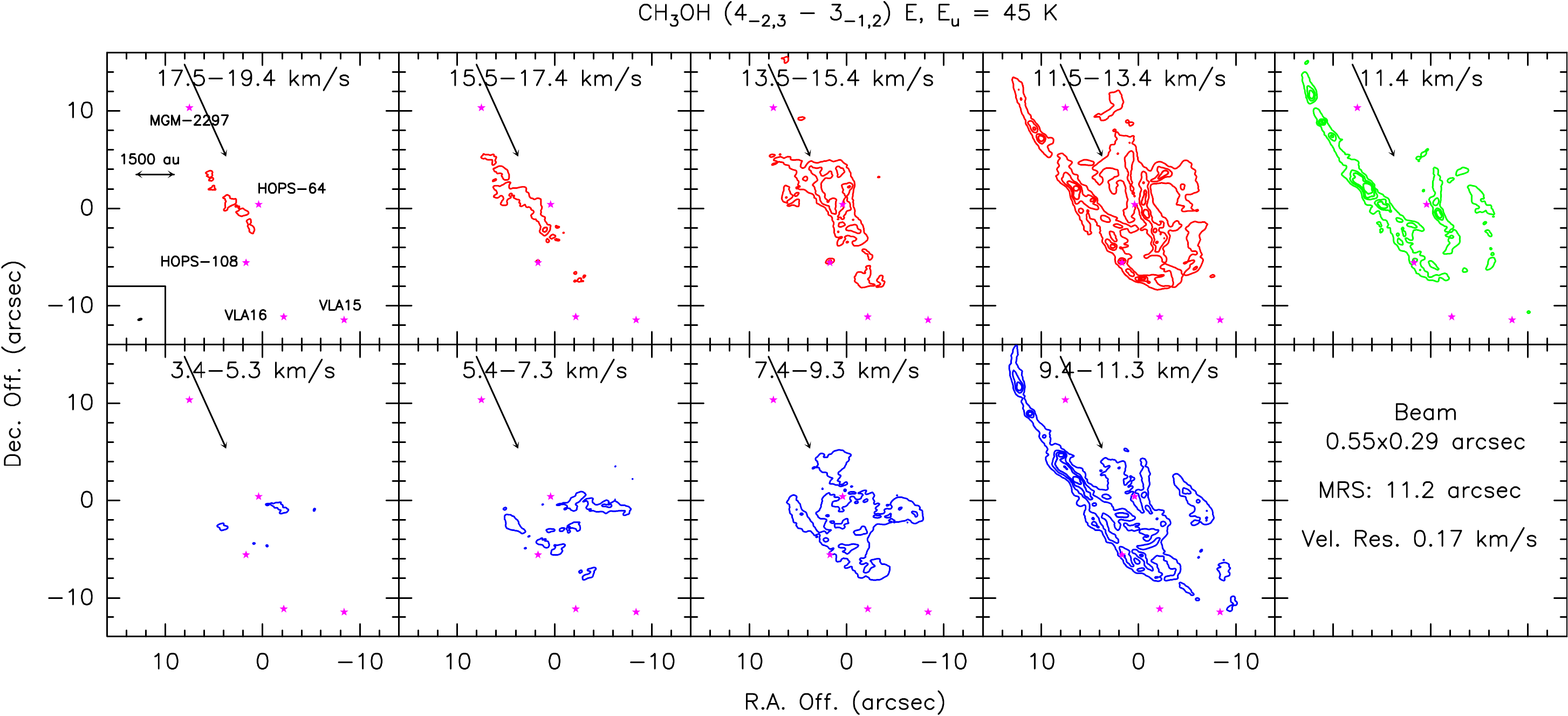}
    \caption{Channel maps of methanol at 50 K (top) and 45 K (bottom). The contours are at [5,20]$\sigma$ (top) and [5, 20, 40, 80]$\sigma$ (bottom) with $\sigma$= (5 mJy beam$^{-1}$\,km\,s$^{-1}$ top, and 8 mJy beam$^{-1}$\,km\,s$^{-1}$ bottom). The cores are depicted with magenta stars and are labelled in black. The synthesised beam is depicted in the lower left corner. We draw one arrow from HOPS-370 perpendicular to its disk. The beam, maximum recoverable angular scale (MRS), and velocity resolution values are reported in the bottom right panel.}
    \label{comp-red-blu-meth}
\end{figure*}

\begin{figure*}[!]
    \centering
    \includegraphics[width=\textwidth]{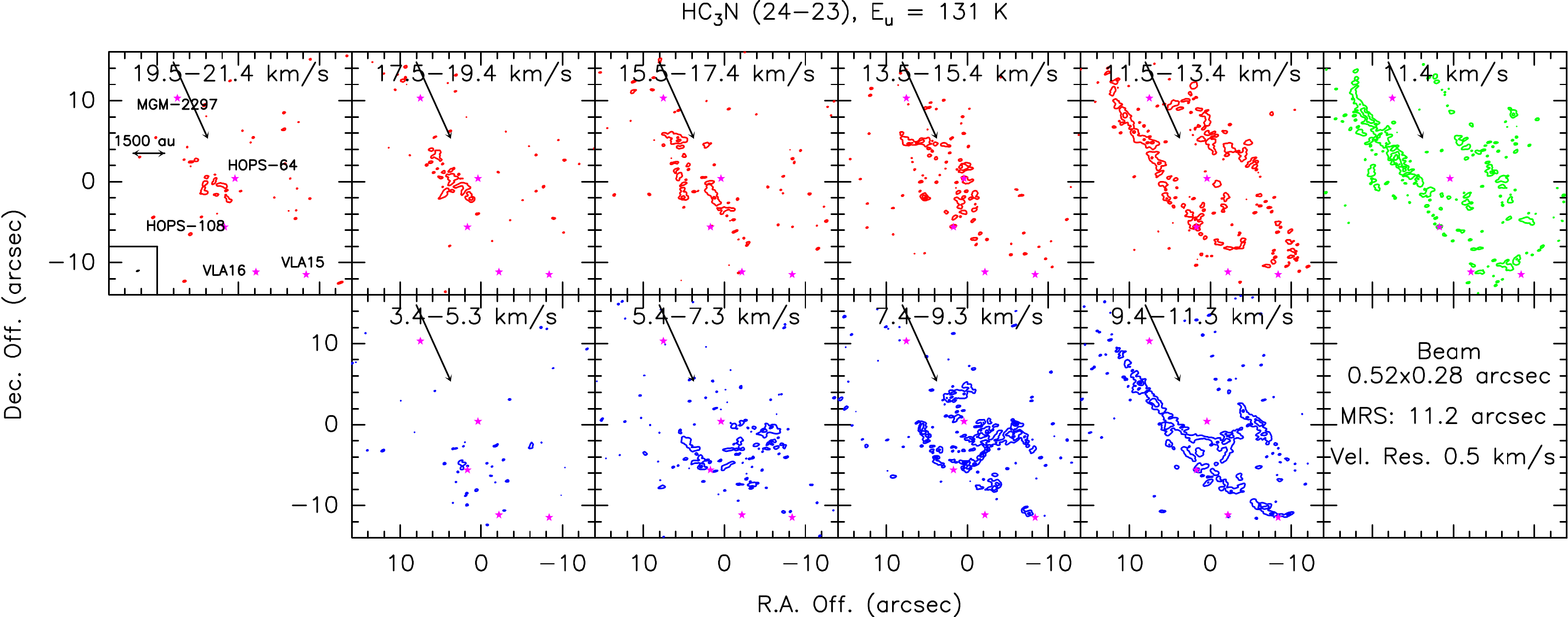}
    \caption{Channel maps of HC$_{3}$N. Contours start at 3$\sigma$ and increase by 5$\sigma$ with $\sigma$= 4 mJy\,beam$^{-1}$\,km\,s$^{-1}$. The cores are depicted with magenta stars and are labelled in black. The synthesised beam is depicted in the lower left corner. We draw one arrow from HOPS-370 perpendicular to its disk. The beam, MRS, and velocity resolution values are reported in the bottom right panel.}
    \label{comp-red-blu-hc3n}
\end{figure*}

\begin{figure*}[!]
    \centering
    \includegraphics[width=0.48\textwidth]{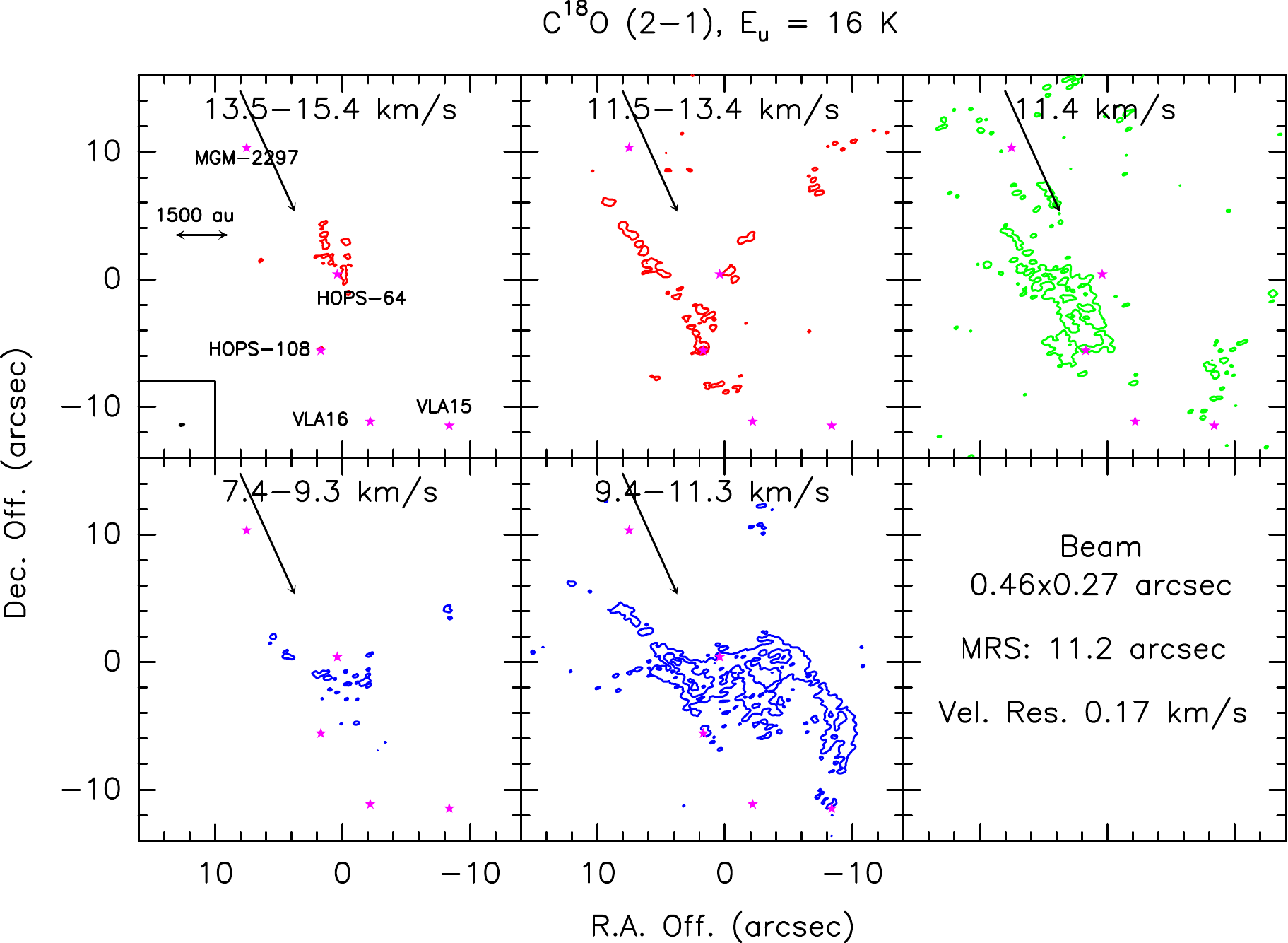}
    \includegraphics[width=0.48\textwidth]{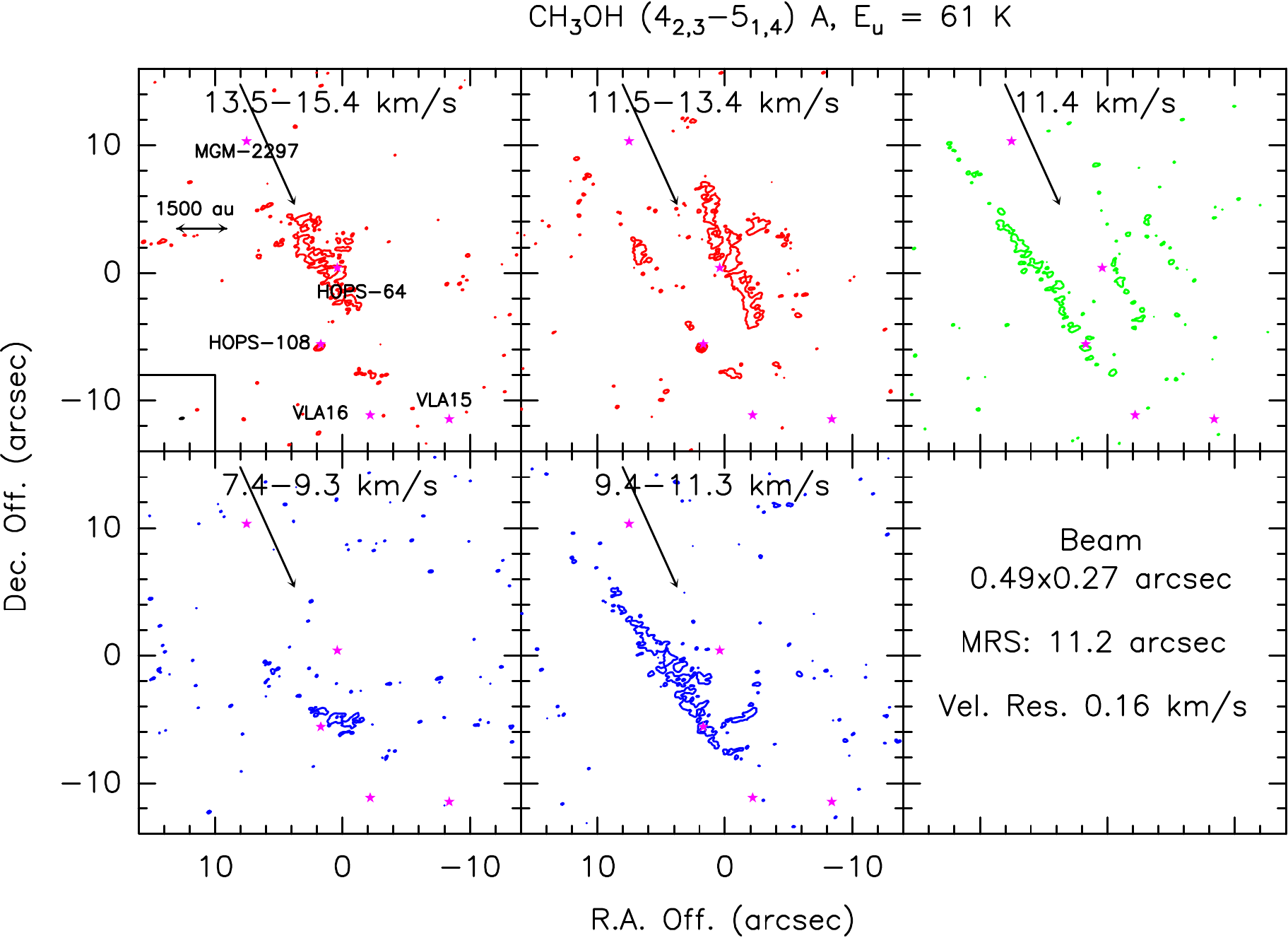}
    \caption{Channel maps of C$^{18}$O (left) and CH$_{3}$OH at 61 K (right). Contours start at 5$\sigma$ for C$^{18}$O, and 3$\sigma$ for CH$_{3}$OH and increase by 5$\sigma$ with $\sigma$= (4 mJy\,beam$^{-1}$\,km\,s$^{-1}$ left, and 6 mJy\,beam$^{-1}$\,km\,s$^{-1}$ right). The cores are depicted with magenta stars and are labelled in black. The synthesised beam is depicted in the lower left corner. We draw one arrow from HOPS-370 perpendicular to its disk. The beam, MRS, and velocity resolution values are reported in the bottom right panel.}
    \label{comp-red-blu-c-18o}
\end{figure*}

\begin{figure*}[!]
    \centering
    \includegraphics[width=\textwidth]{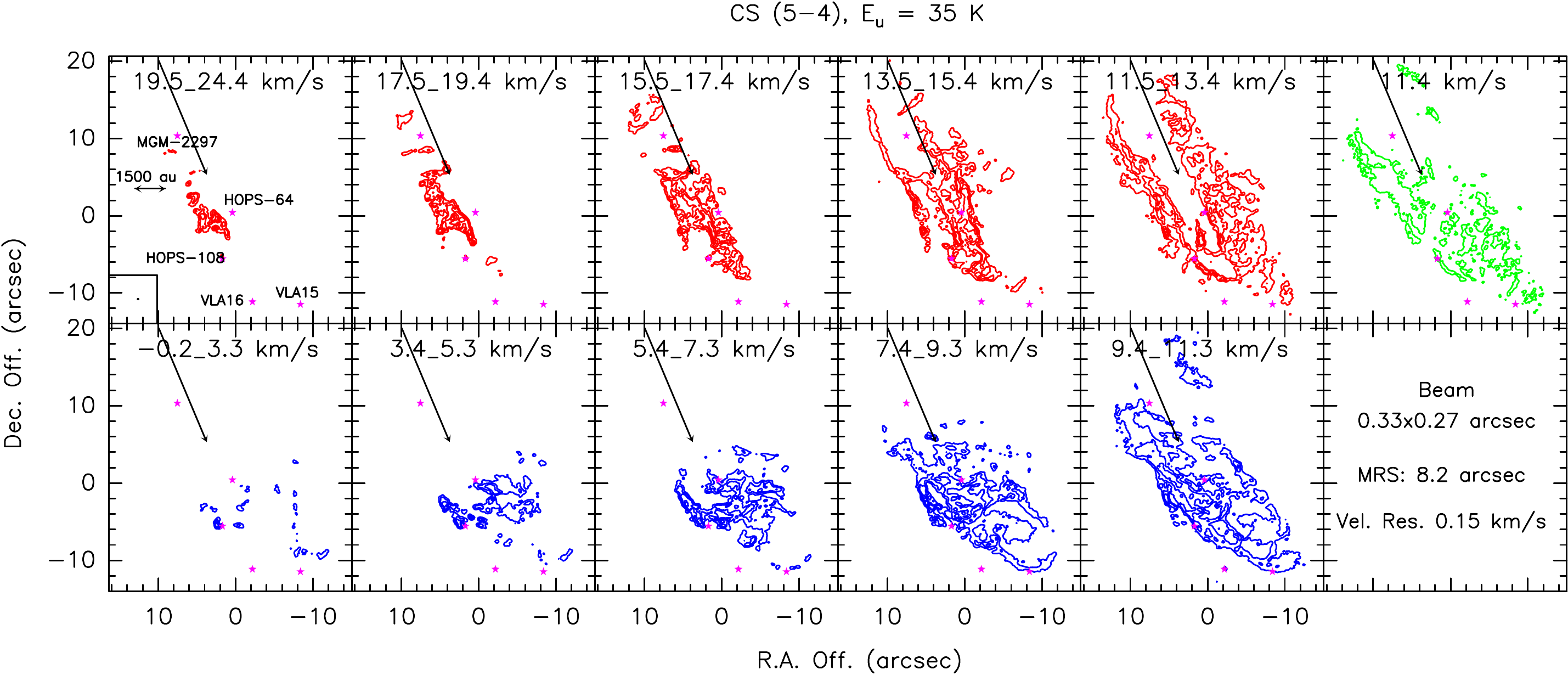}
    \caption{Channel maps of CS. The contours are at [5, 20, 40]$\sigma$ with $\sigma$= 4 mJy\,beam$^{-1}$\,km\,s$^{-1}$. The cores are depicted with magenta stars and are labelled in black. The synthesised beam is depicted in the lower left corner. We draw one arrow from HOPS-370 perpendicular to its disk. The beam, MRS, and velocity resolution values are reported in the bottom right panel.}
    \label{comp-red-blu-cs}
\end{figure*}

\begin{figure*}[!]
    \centering
    \includegraphics[width=\textwidth]{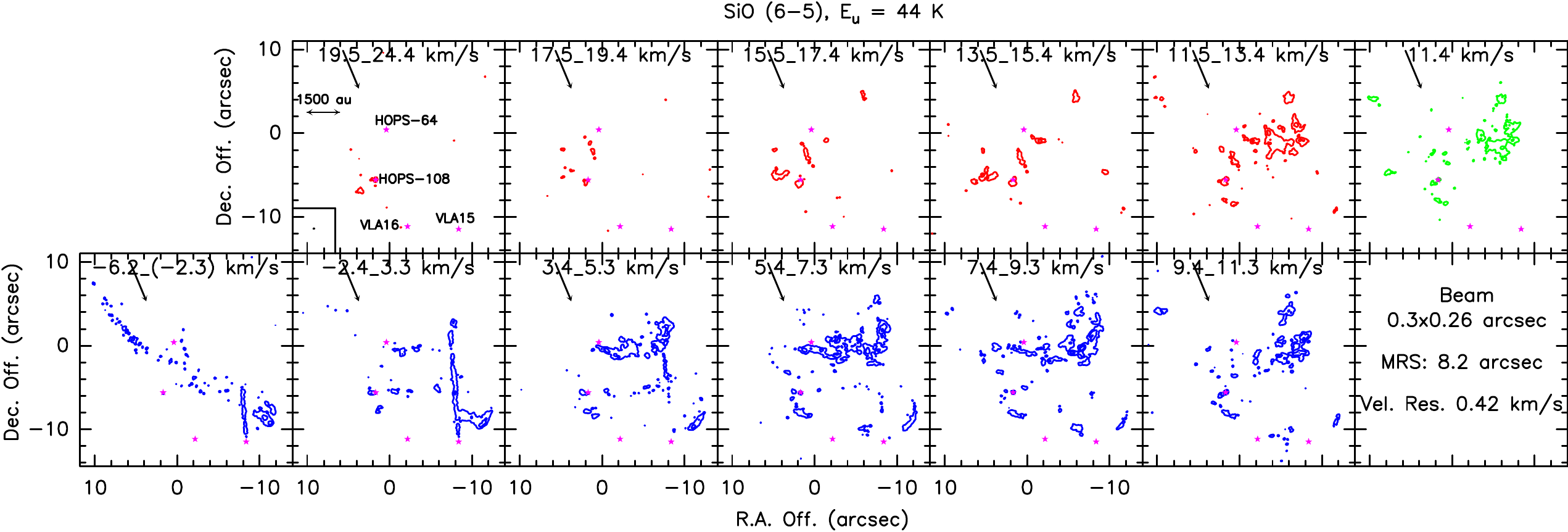}
    \caption{Channel maps of SiO. The contours are at [5, 20, 40]$\sigma$ with $\sigma$= 2.5 mJy\,beam$^{-1}$\,km\,s$^{-1}$. The cores are depicted with magenta stars and are labelled in black. The synthesised beam is depicted in the lower left corner. We draw one arrow from HOPS-370 perpendicular to its disk. The beam, MRS, and velocity resolution values are reported in the bottom right panel.}
    \label{sio-chan-map}
\end{figure*}

In Fig. \ref{sio-continuum} we present the dust emission with superimposed contours of the velocity-integrated emission of SiO at different velocity regimes discussed in the main text. In Figs. \ref{comp-red-blu-meth}-\ref{sio-chan-map} we report the channels maps of CH$_{3}$OH(5$_{1,4}-$4$_{1,3}$\,A), CH$_{3}$OH(4$_{-2,3}-$3$_{-1,2}$\,E), HC$_{3}$N(3--2), C$^{18}$O(2--1), CH$_{3}$OH(4$_{2,3}-$5$_{1,4}$\,E), CS(5--4), and SiO(6--5).

\end{appendix}

\end{document}